\documentclass[aps,prl,reprint,superscriptaddress,longbibliography]{revtex4-1}

\usepackage{amsmath,amssymb}
\usepackage{graphicx}
\usepackage{epstopdf}
\usepackage{bm,ulem}
\usepackage{color}
\usepackage{mhchem}

\newcommand{\fese}     {$\beta$-FeSe}
\newcommand{\bfap}     {BaFe$_2$(As$_{1-x}$P$_x$)$_2$}

\newcommand{\se}	{$^{77}$Se}
\newcommand{\slr} 	{$T_1^{-1}$}
\newcommand{\slrt} 	{$(T_1T)^{-1}$}

\newcommand{\Inmr} 	{$I_\text{NMR}$}
\newcommand{\deltaK} 	{$\Delta \mathcal{K}_{\parallel a}$}

\begin{document}

\title{Nematicity and in-plane anisotropy of
superconductivity in $\beta$-FeSe detected by $^{77}$Se nuclear magnetic resonance}

\author{S.-H. Baek}
\email[]{sbaek.fu@gmail.com}
\affiliation{IFW Dresden, Helmholtzstr. 20, 01069 Dresden, Germany}
\author{D. V. Efremov} 
\affiliation{IFW Dresden, Helmholtzstr. 20, 01069 Dresden, Germany}
\author{J. M. Ok}
\affiliation{Department of Physics, Pohang University of Science and
Technology, Pohang 790-784, Korea}
\author{J. S. Kim}
\affiliation{Department of Physics, Pohang University of Science and
Technology, Pohang 790-784, Korea}
\author{Jeroen van den Brink}
\affiliation{IFW Dresden, Helmholtzstr. 20, 01069 Dresden, Germany}
\affiliation{Department of Physics, Technische Universit\"at Dresden, 01062 Dresden, Germany}
\author{B. B\"uchner}
\affiliation{IFW Dresden, Helmholtzstr. 20, 01069 Dresden, Germany}
\affiliation{Department of Physics, Technische Universit\"at Dresden, 01062 Dresden, Germany}

\date{\today}

\begin{abstract}
	The recent study of $^{77}$Se nuclear magnetic resonance (NMR) in a $\beta$-FeSe single crystal
proposed that
ferro-orbital order breaks the $90^\circ$ $C_4$ rotational symmetry,
driving nematic ordering.
Here, we report an NMR study of the impact of small strains generated by gluing
on nematic state and spin fluctuations.
We observe that the local strains strongly affect the nematic transition, considerably
enhancing its onset temperature. On the contrary, no effect on low-energy spin fluctuations was
found.  
Furthermore we investigate the interplay of the nematic phase and superconductivity.
Our study demonstrates that the twinned
nematic domains respond unequivalently to superconductivity, evidencing the
twofold $C_2$ symmetry of superconductivity in this material.
The obtained results are well understood in terms of the proposed ferro-orbital
order.
\end{abstract}


\maketitle

Many experiments have established the existence
of nematic order --- a state that spontaneously breaks the rotational symmetry
while time-reversal invariance is preserved --- in Fe-based superconductors (FeSCs)
\cite{chuang10,tanatar10a,ying11,song11,yi11,chu12,rosenthal14,iye15}. Although
whether spin or orbital degrees of freedom drive the nematic order is
still under debate \cite{lv09,lv10,liang13,fernandes14,kontani14,chubukov15,yu15},
it is widely believed that a
nematic instability is an important characteristic of the
normal state from which superconductivity emerges. Therefore establishing the
mechanism of the nematic order will help to elucidate
the Cooper pair glue in FeSCs.

In most FeSCs, the nematic state arises in the vicinity of a spin-density
wave state. The temperature ($T$) interval separating them is small.
As a result the strong interaction of various degrees of freedom hides
the nature of the
nematic order. The only exception is \fese\ which has a PbO-type crystal
structure. The nematic order occurs at
$T_{nem} \approx 91$~K and at a lower temperature of $T_c \approx 9$~K
SC sets in. The absence of static magnetism in the whole interval of
temperature together with its simple structure \cite{hsu08} make \fese\ the
primary object for investigation of nematicity and its interplay with
SC \cite{huynh14,nakayama14,shimojima14,chubukov15,
bohmer15,rahn15,wang15a,watson15,mukherjee15,glasbrenner15,rossler15}.
To other spectacular properties of \fese\ belongs the dramatic increase in $T_c$ under
pressure \cite{margadonna09,medvedev09} or by growing mono layer films on
substrates \cite{xiang12,he13,ge15}.

The band structure of \fese\ is typical for FeSC
\cite{maletz14,zhang15}. The low energy is given by two hole
bands around the $\Gamma$ point and two electron bands around the $M$ point. The
nesting between the electron and the hole bands advocates
strong spin fluctuations (SFs) at low $T$ which indeed were observed in
nuclear magnetic resonance (NMR) and neutron studies at
$T\ll T_{nem}$. However no
enhancement of SFs was found close to the nematic
transition. It led to the suggestion that nematic state in \fese\ is driven by orbital
degrees of freedom \cite{wen12b,baek15,yamakawa15,onari15}.

In this Rapid Communication, we report the investigations of nematic order in the normal
and superconducting states of \fese.
We show that gluing the sample introduces random local strains (defects) and
significantly smears out the otherwise sharp nematic transition resulting in
the enhanced onset of nematic ordering. In contrast, it appears that low-energy
SFs are essentially unaffected by the strain.
Furthermore, we demonstrate that the twofold $C_2$ symmetry of
SC in this material is the consequence of the interplay between orbital and
superconducting order parameters.

The \se\ NMR measurements were carried out in a \fese\ single crystal from the
same batch as the one measured in the previous study \cite{baek15}
at an external field ($\mu_0 H$) of 9 T and in the $T$ range of 4.2 -- 180 K.
The sample was mounted to a single-axis goniometer for the exact alignment
along $\mathbf{H}$ within the $a$-$c$ plane of the crystal. The \se\ NMR
spectra were obtained by a standard spin-echo technique, and the \se\
spin-lattice relaxation rate \slr\ was measured using a saturation method by
fitting the recovery of the nuclear magnetization $M(t)$ to a single
exponential function, $1-M(t)/M(\infty)=A\exp(-t/T_1)$, where $A$ is a fitting
parameter that is ideally unity.
In this Rapid Communication, we glued the sample inside the NMR coil using a small amount of
diluted GE varnish not only for the better and stable alignment of the sample,
but also for examining a possible effect of gluing on nematicity.
The motivation is that while using a
glue on samples is unavoidable in many bulk measurements, it is unknown
whether there are nontrivial effects of random strains that could be introduced by
gluing.

Figure 1 (a)
shows \se\ NMR spectra as a function of $T$ at $\mu_0 H=9$
T applied along the crystallographic $a$ axis. The \se\ line splits
into the two lines $\ell_{1,2}$ below $T_{nem}\sim 91$ K, which is consistent
with the previous results \cite{baek15}. In Ref. \cite{baek15}, it was 
established that the 
splitting of the \se\ line is much larger than one could  expect from the 
small orthorhombic distortion.
In order to further demonstrate that the split \se\ NMR lines truly represent the nematic
order parameter, we measured the \se\ spectrum at a fixed $T$ of 60 K as a function
of angle $\theta$ between the $\mathbf{H}$ and the $c$ directions by rotating 
the sample with respect to 
$\mathbf{H}$ from $a$ toward $c$ on the $a$-$c$ plane.   The resultant angle
dependence of the
$\ell_1$-$\ell_2$ spectrum and their Knight shifts $\mathcal{K}$ are shown in
Figs. 1(b) and 1(c), respectively. $\mathcal{K}$'s of
both $\ell_1$ and $\ell_2$ decrease with
decreasing $\theta$ following a cosine function of $\theta$, and smoothly
merge into a single
line when $H\parallel c$. As a result, the difference of the
Knight shifts \deltaK\ [see the
inset of Fig.1(c)] is perfectly described by a cosine function of $\theta$ (solid line).
This unambiguously verifies that \deltaK\ indeed reflects the
nematic order parameter, which is anticipated to obey such a sinusoidal angle
dependence.

\begin{figure}
\centering
\includegraphics[width=\linewidth]{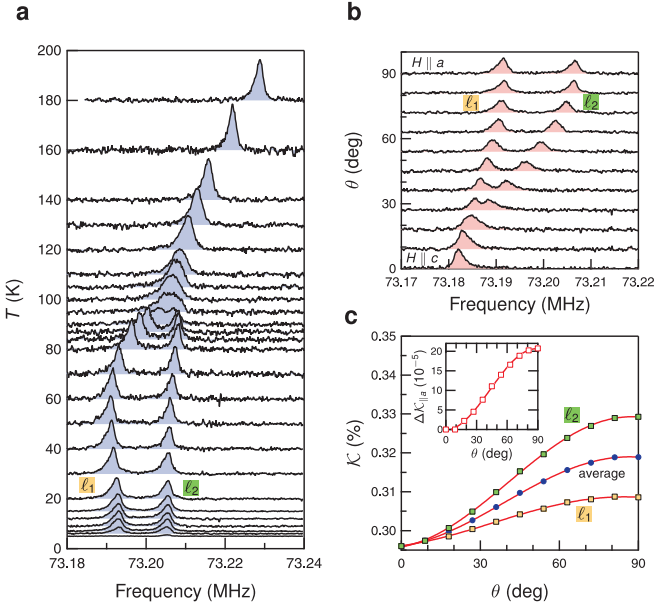}
\caption{(a) The $^{77}$Se NMR spectra for a \fese\ single crystal
measured at $\mu_0 H=9$ T applied along the $a$ axis as a function of
$T$. The splitting of the $^{77}$Se line into two (labeled $\ell_1$
and $\ell_2$) was observed near $T_{nem}\sim 91$ K. (b) A $^{77}$Se spectrum
measured at 60 K as a function of
angle $\theta$ between the $\mathbf{H}$ and the $c$ directions.
(c) Knight shifts of $\ell_1$-$\ell_2$ lines and \deltaK\ (see the
inset) as a function of $\theta$ follow a
simple cosine function (solid lines) as expected for a nematic
order parameter.}
\label{fig:1}
\end{figure}

Although the $T$ evolution of the \se\ NMR spectrum is
very similar to the results obtained in the sample that was not glued
\cite{baek15}, it turns out that
the full width at half maximum of NMR lines are two to three times broader.
Since both the current and the previous NMR studies have been carried out on the
crystals from the same batch under the identical experimental setup and conditions, the
inhomogeneous NMR line broadening is attributed to local strains introduced by
gluing the sample.
In addition to the inhomogeneous broadening of the \se\ lines, we find that
the nematic transition is not as sharp as in the previous study.
Namely, the line splitting does not disappear immediately above the nematic
transition temperature $T_{nem}=91$ K but persists up to higher
$T$.
Indeed, Figs. 2 (a) and 2(b) indicate that the
$\ell_1$ and $\ell_2$ lines start to split at $T^*\sim120$ K that is
significantly higher than $T_{nem}$.
In this regard, we refer to the sample that was not glued as
strain free.
Since in the range of
$T_{nem}<T<T^*$ the material remains in the tetragonal symmetry in bulk,
the persisting NMR line splitting above
$T_{nem}$ indicates that the $90^\circ$ rotational symmetry should be broken
only near local strains.
The reason may be the emergence of the local nematic order $\psi(\mathbf{r})$,
{which exists only locally as a form of small domains without affecting the
bulk tetragonal phase of the crystal,}  by coupling to the
local uniaxial strains 
caused by the gluing.
%
In this case, the broad NMR line between
$T_{nem}\leq T \leq T^*$, which is not well resolved, may
consist of three lines:
the split $\ell_1$-$\ell_2$ lines from the regions near strains
and the unsplit line from the
tetragonal region away from strains.

\begin{figure*}
\centering
\includegraphics[width=\linewidth]{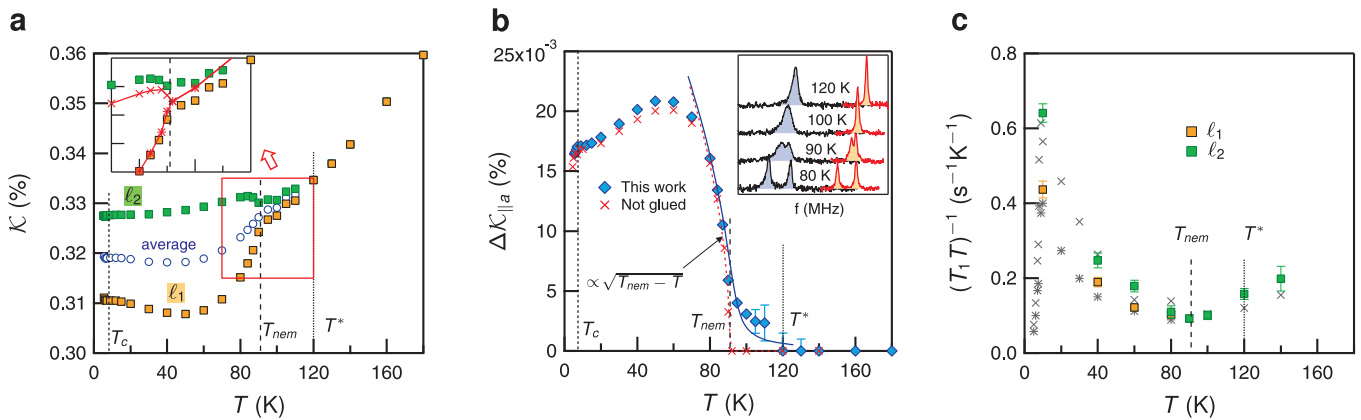}
\caption{(a) and (b) Temperature
dependence of the $^{77}$Se NMR Knight shift $\mathcal{K}$ and the
$\ell_1$-$\ell_2$ splitting \deltaK, respectively. The comparison with the results
in Ref. \cite{baek15} which were rescaled by 1.15 reveals that the actual splitting
occurs at $T^*\sim 120$ K significantly higher than $T_{nem}=91$ K. 
\deltaK\ becomes
proportional to $\sqrt{T_{nem}-T}$ below $T_{nem}$ indicating that $T_{nem}$
is the true nematic transition temperature.
The solid and dashed curves in (b) are theoretical calculations (see the text). The 
inset in (a) shows an enlargement of the region denoted by the red 
rectangle to compare with the sharp splitting observed in a ``strain-free''
sample \cite{baek15}. The raw \se\ spectra from the current Rapid Communication (left) and
those from Ref. \cite{baek15} (right) in selected temperatures near $T_{nem}$
are compared in the inset of (b).
(c) $T$ dependence of
the spin-lattice relaxation rate divided by $T$ \slrt\ near $T_{nem}$. The
asterisk ($\ast$) and cross ($\times$) symbols in gray represent the data
of $\ell_1$ and $\ell_2$, respectively, measured without gluing the sample \cite{baek15}.  }
\label{fig:2}
\end{figure*}

A  description of the $T$ dependence of the nematic order parameter 
$\psi$ can be done in the frame of the Landau theory.  The free energy in 
the presence of an external uniaxial stress $\sigma$ has the following form: 
\begin{equation}
\Delta F = \alpha (T-T_{nem}) \psi^2 + \frac{b}{2} \psi^4 - \lambda  \sigma \psi.
\label{eq.Landau}
\end{equation}
For simplicity we neglect the spatial variation of the nematic order parameter 
assuming that the size of the domains is large. 
The effective 
coupling constant $\lambda$ 
is a $T$ independent function of the shear modulus $C_{66,0}$ that 
can be obtained by integrating out
the structural 
degree of freedom from the Landau functional \cite{bohmer15a}. The linear 
coupling of $\psi$ to $\sigma$ leads to the 
induced nematicity seen in the range of $T_{nem}\leq T\leq T^*$. For 
$T<T_{nem}$, the contribution of $\sigma$ to $\Delta F$ is 
negligible. Therefore the $T$ dependence of $\psi$ is not affected 
by local strains due to gluing, as shown in Fig. 2(b).  The solution $\psi(T)$  
minimizing the Landau functional  
Eq.(\ref{eq.Landau}) is continuous. In the case of a structure with two types of 
domains with strain and without strain, the splitting seen for $T>T_{nem}$ is 
mainly given by the domains with strain, whereas for $T<T_{nem}$ both cases 
give the same NMR line splitting. 
The solutions of Eq. (\ref{eq.Landau}) are given as the solid and 
dashed curves in Fig. 2(b), where the values of the parameters are the same 
for both cases, 
but the solid curve corresponds to 20\%
of the sample volume being under the stress.  
It is interesting to note that
in \bfap\  a similar nematic onset was observed in magnetic torque 
measurements by Kasahara et al.\cite{kasahara12a}.  
For \fese\ we have the advantage to be able to compare strained and unstrained 
samples and observe that at a higher $T$ the onset of the nematic 
ordering is absent in pristine, strain-free crystals, see Fig. 2(b), and only 
present in glued crystals, which are affected by strain. In \fese\ the 
transition at 91 K is therefore of nematic nature and does not share the 
meta-nematic origin that has been assigned to it in \bfap.

Having established that the local strains significantly enhance
local nematicity, the
question arises whether antiferromagnetic SFs are
influenced by them.  To check this, we measured
the spin-lattice relaxation rate divided by $T$ \slrt\ which probes
low-energy SFs.  As shown in Fig. 2(c), \slrt\ is almost
intact near $T_{nem}$ and reveals the same behavior as observed in the
strain-free sample \cite{baek15}. 
This suggests that
the low-energy magnetic excitations are hardly affected by the enhanced 
nematicity.

In parallel to the enhanced onset of nematicity, one can notice that the
nematic order parameter \deltaK\ as
shown in Fig. 2(b) where the previous data were rescaled by 1.15 are 
enhanced compared to the 
previous data. A possible origin is a better alignment of $\mathbf{H}$ along the
$a$ axis in the present Rapid Communication, but it could be also local strains that
enhance the nematicity.
Regardless, an important observation is that \deltaK\ reveals an identical
$T$ dependence with that in Ref. \cite{baek15}. They form
a maximum near 60 K and decrease with decreasing $T$.
At the same time the average of the Knight shifts
$\mathcal{K}_\text{av}=\mathcal{K}_a + \mathcal{K}_b$ and
$\mathcal{K}_{\parallel c}$ show no change with $T$ below 60 K.
The increase in the splitting \deltaK\ in the $T$ interval between
$T_{nem}$ and 60 K can be certainly attributed  to the increase in the
anisotropy of hyperfine couplings
$A^\text{hf}_{xx}-A^\text{hf}_{xx} \propto \psi$ and the
spin susceptibilities $(\chi_{xx}-\chi_{yy}) \propto \psi$ as  \deltaK~
depends on both of these values
$\Delta \mathcal{K}_{\parallel a} =
1/2(A^\text{hf}_{xx} + A^\text{hf}_{yy})(\chi_{xx}-\chi_{yy})+
1/2(A^\text{hf}_{xx} - A^\text{hf}_{yy})(\chi_{xx}+\chi_{yy})$.
The decrease in \deltaK~ with no $T$ change in
$\mathcal{K}_{\parallel c} = A^\text{hf}_c \chi_{zz}$ and the average
$\mathcal{K}_\text{av}=1/2(A^\text{hf}_{xx} + A^\text{hf}_{yy})(\chi_{xx}+\chi_{yy})+
1/2(A^\text{hf}_{xx} - A^\text{hf}_{yy})(\chi_{xx}-\chi_{yy})$
below 60 K is a more subtle question.
It cannot be explained by the nonanalytic contributions to the spin
susceptibility in two-dimensional Fermi liquid as they would also give a linear in $T$
contribution to $\mathcal{K}_{\parallel c}$ and $\mathcal{K}_\text{av}$.
Another interesting observation is that the decrease in \deltaK~ is
accompanied by the increase in the low-energy SFs seen
in the spin lattice relaxation rate \slrt\ [see Fig. 2(c)].
This suggests, contrary to the scenario of the spin nematic
\cite{fernandes14}, that the
increase in SFs results in the suppression of the nematic order.
That is, SFs alone may be insufficient to drive nematic order in
\fese.

\begin{figure}
\centering
\includegraphics[width=0.8\linewidth]{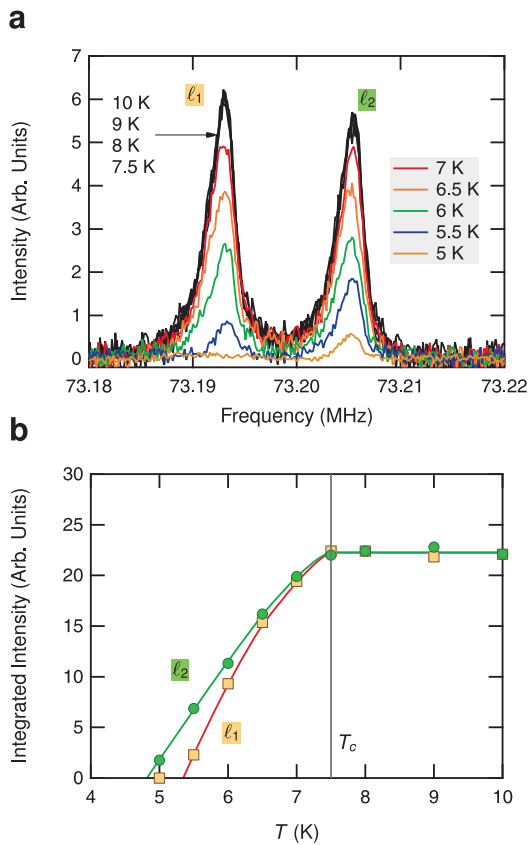}
\caption{(a) Temperature dependence of the split \se\ spectrum measured
at low $T$ for $\mu_0H=9$ T applied along the $a$ axis. Both $\ell_1$ and
$\ell_2$ lines start to lose their intensities below
7.5 K, but the intensity of $\ell_1$ decreases faster than that of
$\ell_2$ with decreasing $T$.
A Boltzmann correction was performed by multiplying $T$ to each spectrum.
(b) Integrated NMR signal intensities of $\ell_1$ and $\ell_2$,
respectively, as a function of $T$. The $\ell_1$ and $\ell_2$ intensities 
are normalized at 10 K.  The different $T$ 
dependence of the $\ell_1$ and
$\ell_2$ signal intensities is clearly revealed, whereas $T_c\sim7.5$ K is
identical for both NMR lines. The solid lines are guides to the eyes.}
\label{fig:3}
\end{figure}

Now we focus on the NMR data in the superconducting state.
In the previous NMR study, we have shown that \deltaK\ abruptly
decreases just below $T_c$.
A similar
anomalous change in \deltaK\ below $T_c$ was
also observed, as shown in Fig. 2 (b), but it appears that the drop of \deltaK\
is much less pronounced than in the previous result. 
This is in line with the consideration that nematicity is strengthened due to 
glueing, rendering it more robust against the competition of the 
superconducting order parameter \cite{bohmer13,baek15}. 
Clearly one expects that the nematic ordering, which appears at a much higher 
$T$ than SC, is only marginally affected by SC, but 
{\it vice versa} that SC is much more strongly affected by the presence 
of already preformed nematic order that breaks the in-plane symmetry. The 
latter we indeed observe as well and is revealed by measurement of  the 
$T$ dependence of intensities of the $\ell_1$ and $\ell_2$ lines below 
$T_c$. 


The $\ell_1$ and $\ell_2$ lines represent the twinned nematic domains,
and their $T$ dependence below $T_c$ is presented in Fig. 2(a).
To begin with, it should be noted that $\ell_1$
has a slightly bigger intensity than $\ell_2$ in the normal state above 7.5 K,
indicating that one of the twinned nematic domains coincidentally has 
larger volume than the 
other. Although the two NMR lines have an unchanged signal intensity
with $T$ in the normal state, we observed that just below $T_c$ they
lose the signal intensity rapidly with decreasing $T$.
In general, the loss of the NMR signal intensity $I_\text{NMR}$ in the SC
state takes place due to diamagnetism of SC.
The penetration of the radiofrequency (rf) pulses into the
bulk sample
is hampered by supercurrents generated on the surface and, as a consequence, the total
number of nuclei that could be irradiated by
the rf pulses (thus \Inmr) is reduced accordingly.
Therefore, one can view that \Inmr\ is proportional to the SC
penetration depth $\lambda$,
although it is difficult to quantify \Inmr\ in terms of $\lambda$ because
there are also many
different contributions to \Inmr\ such as the inhomogeneous field
distribution due to the presence of vortices, vortex dynamics, and the
$Q$-factor of the NMR circuit. 
However, these factors are the same for the two nematic domains that are 
present in the sample -- these domains only differ in their relative 
orientation with respect to the in-plane magnetic field, which is rotated by 
exactly $90^\circ$.

Strikingly, Fig. 3(a) reveals that at 5 K $\ell_1$ is almost quenched
whereas $\ell_2$ is visible, in sharp contrast with the fact that $\ell_1$
has a bigger intensity than $\ell_2$ above $T_c$. This means that $\ell_1$
is losing  intensity faster than $\ell_2$ with lowering $T$ in
the SC state.
Indeed, as shown in Fig. 3(b), the integrated signal intensity as a function
of $T$ reveals the distinctly different
$T$ dependence of $\ell_1$ and $\ell_2$.
Since the split $\ell_1$-$\ell_2$ lines represent the twinned
domains whose nematicity are parallel and perpendicular, respectively, to $\mathbf{H}$, 
without further analysis, we already conclude that the
SC response of the two domains is inequivalent, indicating that
$\lambda_a \neq \lambda_b$. Actually in the clean limit at low $T$ 
the anisotropy in penetration depth
is associated with the anisotropy in the velocities due to the nematic
phase as
$\lambda^{-2}_a/\lambda^{-2}_b \to \langle v_a^2 \rangle/\langle v_b^2 \rangle $
\cite{kogan02}. This implies that the broken in-plane symmetry of
SC is a
natural consequence of the elongated Fermi surface caused by orbital ordering.

This tetragonal symmetry breaking of $\lambda$ is indeed
consistent with the
highly elongated vortex core along the $a$ axis observed by
scanning tunneling microscopy
in FeSe films on a SiC substrate \cite{song11}.
The twofold symmetry of the vortex core state can be
understood by the difference between SC coherence length $\xi$ along the $a$
and $b$ directions, being attributed to the strong impact of the nematic 
order on SC \cite{moon12a,hung12}.
%

%

In conclusion, we presented a NMR study of a \fese\ single crystal under small
strain generated by gluing.
We found that the local strains considerably affect the nematic order, whereas no
effect on low-energy SFs was found. These results suggest that nematicity in \fese\ 
is not driven by SFs, supporting the orbital ordering 
picture.   
The unusual in-plane anisotropy of the penetration depth $\lambda$
manifests the twofold symmetry breaking of SC due to
orbital ordering.

We thank H. Kontani, T. Boothroyd, B. M. Andersen, and R. Fernandes for
useful discussions. This work has been supported
by the Deutsche Forschungsgemeinschaft (Germany) via DFG Research
Grant No. BA 4927/1-3 and the Priority Program SPP 1458.
The work at POSTECH was supported by the NRF through  SRC (Grant No.
2011-0030785) and the Max Planck POSTECH/KOREA Research Initiative (Grant No.
2011-0031558) programs, and by IBS (Grant No. IBSR014-D1-2014-a02) through
the Center for Artificial Low Dimensional Electronic Systems.

\bibliography{mybib}

\begin{thebibliography}{43}%
\makeatletter
\providecommand \@ifxundefined [1]{%
 \@ifx{#1\undefined}
}%
\providecommand \@ifnum [1]{%
 \ifnum #1\expandafter \@firstoftwo
 \else \expandafter \@secondoftwo
 \fi
}%
\providecommand \@ifx [1]{%
 \ifx #1\expandafter \@firstoftwo
 \else \expandafter \@secondoftwo
 \fi
}%
\providecommand \natexlab [1]{#1}%
\providecommand \enquote  [1]{``#1''}%
\providecommand \bibnamefont  [1]{#1}%
\providecommand \bibfnamefont [1]{#1}%
\providecommand \citenamefont [1]{#1}%
\providecommand \href@noop [0]{\@secondoftwo}%
\providecommand \href [0]{\begingroup \@sanitize@url \@href}%
\providecommand \@href[1]{\@@startlink{#1}\@@href}%
\providecommand \@@href[1]{\endgroup#1\@@endlink}%
\providecommand \@sanitize@url [0]{\catcode `\\12\catcode `\$12\catcode
  `\&12\catcode `\#12\catcode `\^12\catcode `\_12\catcode `\%12\relax}%
\providecommand \@@startlink[1]{}%
\providecommand \@@endlink[0]{}%
\providecommand \url  [0]{\begingroup\@sanitize@url \@url }%
\providecommand \@url [1]{\endgroup\@href {#1}{\urlprefix }}%
\providecommand \urlprefix  [0]{URL }%
\providecommand \Eprint [0]{\href }%
\providecommand \doibase [0]{http://dx.doi.org/}%
\providecommand \selectlanguage [0]{\@gobble}%
\providecommand \bibinfo  [0]{\@secondoftwo}%
\providecommand \bibfield  [0]{\@secondoftwo}%
\providecommand \translation [1]{[#1]}%
\providecommand \BibitemOpen [0]{}%
\providecommand \bibitemStop [0]{}%
\providecommand \bibitemNoStop [0]{.\EOS\space}%
\providecommand \EOS [0]{\spacefactor3000\relax}%
\providecommand \BibitemShut  [1]{\csname bibitem#1\endcsname}%
\let\auto@bib@innerbib\@empty
\bibitem [{\citenamefont {Chuang}\ \emph {et~al.}(2010)\citenamefont {Chuang},
  \citenamefont {Allan}, \citenamefont {Lee}, \citenamefont {Xie},
  \citenamefont {Ni}, \citenamefont {Bud'ko}, \citenamefont {Boebinger},
  \citenamefont {Canfield},\ and\ \citenamefont {Davis}}]{chuang10}%
  \BibitemOpen
  \bibfield  {author} {\bibinfo {author} {\bibfnamefont {T.-M.}\ \bibnamefont
  {Chuang}}, \bibinfo {author} {\bibfnamefont {M.~P.}\ \bibnamefont {Allan}},
  \bibinfo {author} {\bibfnamefont {J.}~\bibnamefont {Lee}}, \bibinfo {author}
  {\bibfnamefont {Y.}~\bibnamefont {Xie}}, \bibinfo {author} {\bibfnamefont
  {N.}~\bibnamefont {Ni}}, \bibinfo {author} {\bibfnamefont {S.~L.}\
  \bibnamefont {Bud'ko}}, \bibinfo {author} {\bibfnamefont {G.~S.}\
  \bibnamefont {Boebinger}}, \bibinfo {author} {\bibfnamefont {P.~C.}\
  \bibnamefont {Canfield}}, \ and\ \bibinfo {author} {\bibfnamefont {J.~C.}\
  \bibnamefont {Davis}},\ }\bibfield  {title} {\enquote {\bibinfo {title}
  {{Nematic Electronic Structure in the ``Parent" State of the Iron-Based
  Superconductor Ca(Fe$_{1-x}$Co$_x$)$_2$As$_2$}},}\ }\href {\doibase
  10.1126/science.1181083} {\bibfield  {journal} {\bibinfo  {journal}
  {Science}\ }\textbf {\bibinfo {volume} {327}},\ \bibinfo {pages} {181--184}
  (\bibinfo {year} {2010})}\BibitemShut {NoStop}%
\bibitem [{\citenamefont {Tanatar}\ \emph {et~al.}(2010)\citenamefont
  {Tanatar}, \citenamefont {Blomberg}, \citenamefont {Kreyssig}, \citenamefont
  {Kim}, \citenamefont {Ni}, \citenamefont {Thaler}, \citenamefont {Bud'ko},
  \citenamefont {Canfield}, \citenamefont {Goldman}, \citenamefont {Mazin},\
  and\ \citenamefont {Prozorov}}]{tanatar10a}%
  \BibitemOpen
  \bibfield  {author} {\bibinfo {author} {\bibfnamefont {M.~A.}\ \bibnamefont
  {Tanatar}}, \bibinfo {author} {\bibfnamefont {E.~C.}\ \bibnamefont
  {Blomberg}}, \bibinfo {author} {\bibfnamefont {A.}~\bibnamefont {Kreyssig}},
  \bibinfo {author} {\bibfnamefont {M.~G.}\ \bibnamefont {Kim}}, \bibinfo
  {author} {\bibfnamefont {N.}~\bibnamefont {Ni}}, \bibinfo {author}
  {\bibfnamefont {A.}~\bibnamefont {Thaler}}, \bibinfo {author} {\bibfnamefont
  {S.~L.}\ \bibnamefont {Bud'ko}}, \bibinfo {author} {\bibfnamefont {P.~C.}\
  \bibnamefont {Canfield}}, \bibinfo {author} {\bibfnamefont {A.~I.}\
  \bibnamefont {Goldman}}, \bibinfo {author} {\bibfnamefont {I.~I.}\
  \bibnamefont {Mazin}}, \ and\ \bibinfo {author} {\bibfnamefont
  {R.}~\bibnamefont {Prozorov}},\ }\bibfield  {title} {\enquote {\bibinfo
  {title} {{Uniaxial-strain mechanical detwinning of CaFe$_2$As$_2$ and
  BaFe$_2$As$_2$ crystals: Optical and transport study}},}\ }\href {\doibase
  10.1103/PhysRevB.81.184508} {\bibfield  {journal} {\bibinfo  {journal} {Phys.
  Rev. B}\ }\textbf {\bibinfo {volume} {81}},\ \bibinfo {pages} {184508}
  (\bibinfo {year} {2010})}\BibitemShut {NoStop}%
\bibitem [{\citenamefont {Ying}\ \emph {et~al.}(2011)\citenamefont {Ying},
  \citenamefont {Wang}, \citenamefont {Wu}, \citenamefont {Xiang},
  \citenamefont {Liu}, \citenamefont {Yan}, \citenamefont {Wang}, \citenamefont
  {Zhang}, \citenamefont {Ye}, \citenamefont {Cheng}, \citenamefont {Hu},\ and\
  \citenamefont {Chen}}]{ying11}%
  \BibitemOpen
  \bibfield  {author} {\bibinfo {author} {\bibfnamefont {J.~J.}\ \bibnamefont
  {Ying}}, \bibinfo {author} {\bibfnamefont {X.~F.}\ \bibnamefont {Wang}},
  \bibinfo {author} {\bibfnamefont {T.}~\bibnamefont {Wu}}, \bibinfo {author}
  {\bibfnamefont {Z.~J.}\ \bibnamefont {Xiang}}, \bibinfo {author}
  {\bibfnamefont {R.~H.}\ \bibnamefont {Liu}}, \bibinfo {author} {\bibfnamefont
  {Y.~J.}\ \bibnamefont {Yan}}, \bibinfo {author} {\bibfnamefont {A.~F.}\
  \bibnamefont {Wang}}, \bibinfo {author} {\bibfnamefont {M.}~\bibnamefont
  {Zhang}}, \bibinfo {author} {\bibfnamefont {G.~J.}\ \bibnamefont {Ye}},
  \bibinfo {author} {\bibfnamefont {P.}~\bibnamefont {Cheng}}, \bibinfo
  {author} {\bibfnamefont {J.~P.}\ \bibnamefont {Hu}}, \ and\ \bibinfo {author}
  {\bibfnamefont {X.~H.}\ \bibnamefont {Chen}},\ }\bibfield  {title} {\enquote
  {\bibinfo {title} {Measurements of the anisotropic in-plane resistivity of
  underdoped feas-based pnictide superconductors},}\ }\href {\doibase
  10.1103/PhysRevLett.107.067001} {\bibfield  {journal} {\bibinfo  {journal}
  {Phys. Rev. Lett.}\ }\textbf {\bibinfo {volume} {107}},\ \bibinfo {pages}
  {067001} (\bibinfo {year} {2011})}\BibitemShut {NoStop}%
\bibitem [{\citenamefont {Song}\ \emph {et~al.}(2011)\citenamefont {Song},
  \citenamefont {Wang}, \citenamefont {Cheng}, \citenamefont {Jiang},
  \citenamefont {Li}, \citenamefont {Zhang}, \citenamefont {Li}, \citenamefont
  {He}, \citenamefont {Wang}, \citenamefont {Jia}, \citenamefont {Hung},
  \citenamefont {Wu}, \citenamefont {Ma}, \citenamefont {Chen},\ and\
  \citenamefont {Xue}}]{song11}%
  \BibitemOpen
  \bibfield  {author} {\bibinfo {author} {\bibfnamefont {C.-L.}\ \bibnamefont
  {Song}}, \bibinfo {author} {\bibfnamefont {Y.-L.}\ \bibnamefont {Wang}},
  \bibinfo {author} {\bibfnamefont {P.}~\bibnamefont {Cheng}}, \bibinfo
  {author} {\bibfnamefont {Y.-P.}\ \bibnamefont {Jiang}}, \bibinfo {author}
  {\bibfnamefont {W.}~\bibnamefont {Li}}, \bibinfo {author} {\bibfnamefont
  {T.}~\bibnamefont {Zhang}}, \bibinfo {author} {\bibfnamefont
  {Z.}~\bibnamefont {Li}}, \bibinfo {author} {\bibfnamefont {K.}~\bibnamefont
  {He}}, \bibinfo {author} {\bibfnamefont {L.}~\bibnamefont {Wang}}, \bibinfo
  {author} {\bibfnamefont {J.-F.}\ \bibnamefont {Jia}}, \bibinfo {author}
  {\bibfnamefont {H.-H.}\ \bibnamefont {Hung}}, \bibinfo {author}
  {\bibfnamefont {C.}~\bibnamefont {Wu}}, \bibinfo {author} {\bibfnamefont
  {X.}~\bibnamefont {Ma}}, \bibinfo {author} {\bibfnamefont {X.}~\bibnamefont
  {Chen}}, \ and\ \bibinfo {author} {\bibfnamefont {Q.-K.}\ \bibnamefont
  {Xue}},\ }\bibfield  {title} {\enquote {\bibinfo {title} {{Direct Observation
  of Nodes and Twofold Symmetry in FeSe Superconductor}},}\ }\href {\doibase
  10.1126/science.1202226} {\bibfield  {journal} {\bibinfo  {journal}
  {Science}\ }\textbf {\bibinfo {volume} {332}},\ \bibinfo {pages} {1410--1413}
  (\bibinfo {year} {2011})}\BibitemShut {NoStop}%
\bibitem [{\citenamefont {Yi}\ \emph {et~al.}(2011)\citenamefont {Yi},
  \citenamefont {Lu}, \citenamefont {Chu}, \citenamefont {Analytis},
  \citenamefont {Sorini}, \citenamefont {Kemper}, \citenamefont {Moritz},
  \citenamefont {Mo}, \citenamefont {Moore}, \citenamefont {Hashimoto},
  \citenamefont {Lee}, \citenamefont {Hussain}, \citenamefont {Devereaux},
  \citenamefont {Fisher},\ and\ \citenamefont {Shen}}]{yi11}%
  \BibitemOpen
  \bibfield  {author} {\bibinfo {author} {\bibfnamefont {M.}~\bibnamefont
  {Yi}}, \bibinfo {author} {\bibfnamefont {D.}~\bibnamefont {Lu}}, \bibinfo
  {author} {\bibfnamefont {J.-H.}\ \bibnamefont {Chu}}, \bibinfo {author}
  {\bibfnamefont {J.~G.}\ \bibnamefont {Analytis}}, \bibinfo {author}
  {\bibfnamefont {A.~P.}\ \bibnamefont {Sorini}}, \bibinfo {author}
  {\bibfnamefont {A.~F.}\ \bibnamefont {Kemper}}, \bibinfo {author}
  {\bibfnamefont {B.}~\bibnamefont {Moritz}}, \bibinfo {author} {\bibfnamefont
  {S.-K.}\ \bibnamefont {Mo}}, \bibinfo {author} {\bibfnamefont {R.~G.}\
  \bibnamefont {Moore}}, \bibinfo {author} {\bibfnamefont {M.}~\bibnamefont
  {Hashimoto}}, \bibinfo {author} {\bibfnamefont {W.-S.}\ \bibnamefont {Lee}},
  \bibinfo {author} {\bibfnamefont {Z.}~\bibnamefont {Hussain}}, \bibinfo
  {author} {\bibfnamefont {T.~P.}\ \bibnamefont {Devereaux}}, \bibinfo {author}
  {\bibfnamefont {I.~R.}\ \bibnamefont {Fisher}}, \ and\ \bibinfo {author}
  {\bibfnamefont {Z.-X.}\ \bibnamefont {Shen}},\ }\bibfield  {title} {\enquote
  {\bibinfo {title} {{Symmetry-breaking orbital anisotropy observed for
  detwinned Ba(Fe$_{1-x}$Co$_x$)$_2$As$_2$ above the spin density wave
  transition}},}\ }\href {\doibase 10.1073/pnas.1015572108} {\bibfield
  {journal} {\bibinfo  {journal} {Proc. Natl. Acad. Sci. U.S.A.}\ }\textbf
  {\bibinfo {volume} {108}},\ \bibinfo {pages} {6878--6883} (\bibinfo {year}
  {2011})}\BibitemShut {NoStop}%
\bibitem [{\citenamefont {Chu}\ \emph {et~al.}(2012)\citenamefont {Chu},
  \citenamefont {Kuo}, \citenamefont {Analytis},\ and\ \citenamefont
  {Fisher}}]{chu12}%
  \BibitemOpen
  \bibfield  {author} {\bibinfo {author} {\bibfnamefont {J.-H.}\ \bibnamefont
  {Chu}}, \bibinfo {author} {\bibfnamefont {H.-H.}\ \bibnamefont {Kuo}},
  \bibinfo {author} {\bibfnamefont {J.~G.}\ \bibnamefont {Analytis}}, \ and\
  \bibinfo {author} {\bibfnamefont {I.~R.}\ \bibnamefont {Fisher}},\ }\bibfield
   {title} {\enquote {\bibinfo {title} {Divergent nematic susceptibility in an
  iron arsenide superconductor},}\ }\href {\doibase 10.1126/science.1221713}
  {\bibfield  {journal} {\bibinfo  {journal} {Science}\ }\textbf {\bibinfo
  {volume} {337}},\ \bibinfo {pages} {710--712} (\bibinfo {year}
  {2012})}\BibitemShut {NoStop}%
\bibitem [{\citenamefont {Rosenthal}\ \emph {et~al.}(2014)\citenamefont
  {Rosenthal}, \citenamefont {Andrade}, \citenamefont {Arguello}, \citenamefont
  {Fernandes}, \citenamefont {Xing}, \citenamefont {Wang}, \citenamefont {Jin},
  \citenamefont {Millis},\ and\ \citenamefont {Pasupathy}}]{rosenthal14}%
  \BibitemOpen
  \bibfield  {author} {\bibinfo {author} {\bibfnamefont {E.~P.}\ \bibnamefont
  {Rosenthal}}, \bibinfo {author} {\bibfnamefont {E.~F.}\ \bibnamefont
  {Andrade}}, \bibinfo {author} {\bibfnamefont {C.~J.}\ \bibnamefont
  {Arguello}}, \bibinfo {author} {\bibfnamefont {R.~M.}\ \bibnamefont
  {Fernandes}}, \bibinfo {author} {\bibfnamefont {L.~Y.}\ \bibnamefont {Xing}},
  \bibinfo {author} {\bibfnamefont {X.~C.}\ \bibnamefont {Wang}}, \bibinfo
  {author} {\bibfnamefont {C.~Q.}\ \bibnamefont {Jin}}, \bibinfo {author}
  {\bibfnamefont {A.~J.}\ \bibnamefont {Millis}}, \ and\ \bibinfo {author}
  {\bibfnamefont {A.~N.}\ \bibnamefont {Pasupathy}},\ }\bibfield  {title}
  {\enquote {\bibinfo {title} {{Visualization of electron nematicity and
  unidirectional antiferroic fluctuations at high temperatures in NaFeAs}},}\
  }\href {\doibase 10.1038/nphys2870} {\bibfield  {journal} {\bibinfo
  {journal} {Nature Phys.}\ }\textbf {\bibinfo {volume} {10}},\ \bibinfo
  {pages} {225--232} (\bibinfo {year} {2014})}\BibitemShut {NoStop}%
\bibitem [{\citenamefont {Iye}\ \emph {et~al.}(2015)\citenamefont {Iye},
  \citenamefont {Julien}, \citenamefont {Mayaffre}, \citenamefont
  {Horvati\'c.}, \citenamefont {Berthier}, \citenamefont {Ishida},
  \citenamefont {Ikeda}, \citenamefont {Kasahara}, \citenamefont {Shibauchi},\
  and\ \citenamefont {Matsuda}}]{iye15}%
  \BibitemOpen
  \bibfield  {author} {\bibinfo {author} {\bibfnamefont {T.}~\bibnamefont
  {Iye}}, \bibinfo {author} {\bibfnamefont {M.-H.}\ \bibnamefont {Julien}},
  \bibinfo {author} {\bibfnamefont {H.}~\bibnamefont {Mayaffre}}, \bibinfo
  {author} {\bibfnamefont {M.}~\bibnamefont {Horvati\'c.}}, \bibinfo {author}
  {\bibfnamefont {C.}~\bibnamefont {Berthier}}, \bibinfo {author}
  {\bibfnamefont {K.}~\bibnamefont {Ishida}}, \bibinfo {author} {\bibfnamefont
  {H.}~\bibnamefont {Ikeda}}, \bibinfo {author} {\bibfnamefont
  {S.}~\bibnamefont {Kasahara}}, \bibinfo {author} {\bibfnamefont
  {T.}~\bibnamefont {Shibauchi}}, \ and\ \bibinfo {author} {\bibfnamefont
  {Y.}~\bibnamefont {Matsuda}},\ }\bibfield  {title} {\enquote {\bibinfo
  {title} {{Emergence of Orbital Nematicity in the Tetragonal Phase of
  BaFe$_2$(As$_{1-x}$P$_{x}$)$_2$}},}\ }\href {\doibase 10.7566/JPSJ.84.043705}
  {\bibfield  {journal} {\bibinfo  {journal} {J. Phys. Soc. Jpn.}\ }\textbf
  {\bibinfo {volume} {84}},\ \bibinfo {pages} {043705} (\bibinfo {year}
  {2015})}\BibitemShut {NoStop}%
\bibitem [{\citenamefont {Lv}\ \emph {et~al.}(2009)\citenamefont {Lv},
  \citenamefont {Wu},\ and\ \citenamefont {Phillips}}]{lv09}%
  \BibitemOpen
  \bibfield  {author} {\bibinfo {author} {\bibfnamefont {W.}~\bibnamefont
  {Lv}}, \bibinfo {author} {\bibfnamefont {J.}~\bibnamefont {Wu}}, \ and\
  \bibinfo {author} {\bibfnamefont {P.}~\bibnamefont {Phillips}},\ }\bibfield
  {title} {\enquote {\bibinfo {title} {{Orbital ordering induces structural
  phase transition and the resistivity anomaly in iron pnictides}},}\ }\href
  {\doibase 10.1103/PhysRevB.80.224506} {\bibfield  {journal} {\bibinfo
  {journal} {Phys. Rev. B}\ }\textbf {\bibinfo {volume} {80}},\ \bibinfo
  {pages} {224506} (\bibinfo {year} {2009})}\BibitemShut {NoStop}%
\bibitem [{\citenamefont {Lv}\ \emph {et~al.}(2010)\citenamefont {Lv},
  \citenamefont {Kr\"uger},\ and\ \citenamefont {Phillips}}]{lv10}%
  \BibitemOpen
  \bibfield  {author} {\bibinfo {author} {\bibfnamefont {W.}~\bibnamefont
  {Lv}}, \bibinfo {author} {\bibfnamefont {F.}~\bibnamefont {Kr\"uger}}, \ and\
  \bibinfo {author} {\bibfnamefont {P.}~\bibnamefont {Phillips}},\ }\bibfield
  {title} {\enquote {\bibinfo {title} {{Orbital ordering and unfrustrated
  $(\pi,0)$ magnetism from degenerate double exchange in the iron
  pnictides}},}\ }\href {\doibase 10.1103/PhysRevB.82.045125} {\bibfield
  {journal} {\bibinfo  {journal} {Phys. Rev. B}\ }\textbf {\bibinfo {volume}
  {82}},\ \bibinfo {pages} {045125} (\bibinfo {year} {2010})}\BibitemShut
  {NoStop}%
\bibitem [{\citenamefont {Liang}\ \emph {et~al.}(2013)\citenamefont {Liang},
  \citenamefont {Moreo},\ and\ \citenamefont {Dagotto}}]{liang13}%
  \BibitemOpen
  \bibfield  {author} {\bibinfo {author} {\bibfnamefont {S.}~\bibnamefont
  {Liang}}, \bibinfo {author} {\bibfnamefont {A.}~\bibnamefont {Moreo}}, \ and\
  \bibinfo {author} {\bibfnamefont {E.}~\bibnamefont {Dagotto}},\ }\bibfield
  {title} {\enquote {\bibinfo {title} {{Nematic State of Pnictides Stabilized
  by Interplay between Spin, Orbital, and Lattice Degrees of Freedom}},}\
  }\href {\doibase 10.1103/PhysRevLett.111.047004} {\bibfield  {journal}
  {\bibinfo  {journal} {Phys. Rev. Lett.}\ }\textbf {\bibinfo {volume} {111}},\
  \bibinfo {pages} {047004} (\bibinfo {year} {2013})}\BibitemShut {NoStop}%
\bibitem [{\citenamefont {Fernandes}\ \emph {et~al.}(2014)\citenamefont
  {Fernandes}, \citenamefont {Chubukov},\ and\ \citenamefont
  {Schmalian}}]{fernandes14}%
  \BibitemOpen
  \bibfield  {author} {\bibinfo {author} {\bibfnamefont {R.~M.}\ \bibnamefont
  {Fernandes}}, \bibinfo {author} {\bibfnamefont {A.~V.}\ \bibnamefont
  {Chubukov}}, \ and\ \bibinfo {author} {\bibfnamefont {J.}~\bibnamefont
  {Schmalian}},\ }\bibfield  {title} {\enquote {\bibinfo {title} {What drives
  nematic order in iron-based superconductors?}}\ }\href {\doibase
  10.1038/nphys2877} {\bibfield  {journal} {\bibinfo  {journal} {Nature Phys.}\
  }\textbf {\bibinfo {volume} {10}},\ \bibinfo {pages} {97--104} (\bibinfo
  {year} {2014})}\BibitemShut {NoStop}%
\bibitem [{\citenamefont {Kontani}\ and\ \citenamefont
  {Yamakawa}(2014)}]{kontani14}%
  \BibitemOpen
  \bibfield  {author} {\bibinfo {author} {\bibfnamefont {H.}~\bibnamefont
  {Kontani}}\ and\ \bibinfo {author} {\bibfnamefont {Y.}~\bibnamefont
  {Yamakawa}},\ }\bibfield  {title} {\enquote {\bibinfo {title} {{Linear
  Response Theory for Shear Modulus ${C}_{66}$ and Raman Quadrupole
  Susceptibility: Evidence for Nematic Orbital Fluctuations in Fe-based
  Superconductors}},}\ }\href {\doibase 10.1103/PhysRevLett.113.047001}
  {\bibfield  {journal} {\bibinfo  {journal} {Phys. Rev. Lett.}\ }\textbf
  {\bibinfo {volume} {113}},\ \bibinfo {pages} {047001} (\bibinfo {year}
  {2014})}\BibitemShut {NoStop}%
\bibitem [{\citenamefont {Chubukov}\ \emph {et~al.}(2015)\citenamefont
  {Chubukov}, \citenamefont {Fernandes},\ and\ \citenamefont
  {Schmalian}}]{chubukov15}%
  \BibitemOpen
  \bibfield  {author} {\bibinfo {author} {\bibfnamefont {A.~V.}\ \bibnamefont
  {Chubukov}}, \bibinfo {author} {\bibfnamefont {R.~M.}\ \bibnamefont
  {Fernandes}}, \ and\ \bibinfo {author} {\bibfnamefont {J.}~\bibnamefont
  {Schmalian}},\ }\bibfield  {title} {\enquote {\bibinfo {title} {{Origin of
  nematic order in FeSe}},}\ }\href {\doibase 10.1103/PhysRevB.91.201105}
  {\bibfield  {journal} {\bibinfo  {journal} {Phys. Rev. B}\ }\textbf {\bibinfo
  {volume} {91}},\ \bibinfo {pages} {201105} (\bibinfo {year}
  {2015})}\BibitemShut {NoStop}%
\bibitem [{\citenamefont {Yu}\ and\ \citenamefont {Si}(2015)}]{yu15}%
  \BibitemOpen
  \bibfield  {author} {\bibinfo {author} {\bibfnamefont {R.}~\bibnamefont
  {Yu}}\ and\ \bibinfo {author} {\bibfnamefont {Q.}~\bibnamefont {Si}},\
  }\bibfield  {title} {\enquote {\bibinfo {title} {{Antiferroquadrupolar and
  Ising-Nematic Orders of a Frustrated Bilinear-Biquadratic Heisenberg Model
  and Implications for the Magnetism of FeSe}},}\ }\href {\doibase
  10.1103/PhysRevLett.115.116401} {\bibfield  {journal} {\bibinfo  {journal}
  {Phys. Rev. Lett.}\ }\textbf {\bibinfo {volume} {115}},\ \bibinfo {pages}
  {116401} (\bibinfo {year} {2015})}\BibitemShut {NoStop}%
\bibitem [{\citenamefont {Hsu}\ \emph {et~al.}(2008)\citenamefont {Hsu},
  \citenamefont {Luo}, \citenamefont {Yeh}, \citenamefont {Chen}, \citenamefont
  {Huang}, \citenamefont {Wu}, \citenamefont {Lee}, \citenamefont {Huang},
  \citenamefont {Chu}, \citenamefont {Yan},\ and\ \citenamefont {Wu}}]{hsu08}%
  \BibitemOpen
  \bibfield  {author} {\bibinfo {author} {\bibfnamefont {F.-C.}\ \bibnamefont
  {Hsu}}, \bibinfo {author} {\bibfnamefont {J.-Y.}\ \bibnamefont {Luo}},
  \bibinfo {author} {\bibfnamefont {K.-W.}\ \bibnamefont {Yeh}}, \bibinfo
  {author} {\bibfnamefont {T.-K.}\ \bibnamefont {Chen}}, \bibinfo {author}
  {\bibfnamefont {T.-W.}\ \bibnamefont {Huang}}, \bibinfo {author}
  {\bibfnamefont {P.~M.}\ \bibnamefont {Wu}}, \bibinfo {author} {\bibfnamefont
  {Y.-C.}\ \bibnamefont {Lee}}, \bibinfo {author} {\bibfnamefont {Y.-L.}\
  \bibnamefont {Huang}}, \bibinfo {author} {\bibfnamefont {Y.-Y.}\ \bibnamefont
  {Chu}}, \bibinfo {author} {\bibfnamefont {D.-C.}\ \bibnamefont {Yan}}, \ and\
  \bibinfo {author} {\bibfnamefont {M.-K.}\ \bibnamefont {Wu}},\ }\bibfield
  {title} {\enquote {\bibinfo {title} {{Superconductivity in the {PbO}-type
  structure {$\alpha$-FeSe}}},}\ }\href {\doibase 10.1073/pnas.0807325105}
  {\bibfield  {journal} {\bibinfo  {journal} {Proc. Nat. Acad. Sci.}\ }\textbf
  {\bibinfo {volume} {105}},\ \bibinfo {pages} {14262--14264} (\bibinfo {year}
  {2008})}\BibitemShut {NoStop}%
\bibitem [{\citenamefont {Huynh}\ \emph {et~al.}(2014)\citenamefont {Huynh},
  \citenamefont {Tanabe}, \citenamefont {Urata}, \citenamefont {Oguro},
  \citenamefont {Heguri}, \citenamefont {Watanabe},\ and\ \citenamefont
  {Tanigaki}}]{huynh14}%
  \BibitemOpen
  \bibfield  {author} {\bibinfo {author} {\bibfnamefont {K.~K.}\ \bibnamefont
  {Huynh}}, \bibinfo {author} {\bibfnamefont {Y.}~\bibnamefont {Tanabe}},
  \bibinfo {author} {\bibfnamefont {T.}~\bibnamefont {Urata}}, \bibinfo
  {author} {\bibfnamefont {H.}~\bibnamefont {Oguro}}, \bibinfo {author}
  {\bibfnamefont {S.}~\bibnamefont {Heguri}}, \bibinfo {author} {\bibfnamefont
  {K.}~\bibnamefont {Watanabe}}, \ and\ \bibinfo {author} {\bibfnamefont
  {K.}~\bibnamefont {Tanigaki}},\ }\bibfield  {title} {\enquote {\bibinfo
  {title} {{Electric transport of a single-crystal iron chalcogenide FeSe
  superconductor: Evidence of symmetry-breakdown nematicity and additional
  ultrafast Dirac cone-like carriers}},}\ }\href {\doibase
  10.1103/PhysRevB.90.144516} {\bibfield  {journal} {\bibinfo  {journal} {Phys.
  Rev. B}\ }\textbf {\bibinfo {volume} {90}},\ \bibinfo {pages} {144516}
  (\bibinfo {year} {2014})}\BibitemShut {NoStop}%
\bibitem [{\citenamefont {Nakayama}\ \emph {et~al.}(2014)\citenamefont
  {Nakayama}, \citenamefont {Miyata}, \citenamefont {Phan}, \citenamefont
  {Sato}, \citenamefont {Tanabe}, \citenamefont {Urata}, \citenamefont
  {Tanigaki},\ and\ \citenamefont {Takahashi}}]{nakayama14}%
  \BibitemOpen
  \bibfield  {author} {\bibinfo {author} {\bibfnamefont {K.}~\bibnamefont
  {Nakayama}}, \bibinfo {author} {\bibfnamefont {Y.}~\bibnamefont {Miyata}},
  \bibinfo {author} {\bibfnamefont {G.~N.}\ \bibnamefont {Phan}}, \bibinfo
  {author} {\bibfnamefont {T.}~\bibnamefont {Sato}}, \bibinfo {author}
  {\bibfnamefont {Y.}~\bibnamefont {Tanabe}}, \bibinfo {author} {\bibfnamefont
  {T.}~\bibnamefont {Urata}}, \bibinfo {author} {\bibfnamefont
  {K.}~\bibnamefont {Tanigaki}}, \ and\ \bibinfo {author} {\bibfnamefont
  {T.}~\bibnamefont {Takahashi}},\ }\bibfield  {title} {\enquote {\bibinfo
  {title} {{Reconstruction of Band Structure Induced by Electronic Nematicity
  in an FeSe Superconductor}},}\ }\href {\doibase
  10.1103/PhysRevLett.113.237001} {\bibfield  {journal} {\bibinfo  {journal}
  {Phys. Rev. Lett.}\ }\textbf {\bibinfo {volume} {113}},\ \bibinfo {pages}
  {237001} (\bibinfo {year} {2014})}\BibitemShut {NoStop}%
\bibitem [{\citenamefont {Shimojima}\ \emph {et~al.}(2014)\citenamefont
  {Shimojima}, \citenamefont {Suzuki}, \citenamefont {Sonobe}, \citenamefont
  {Nakamura}, \citenamefont {Sakano}, \citenamefont {Omachi}, \citenamefont
  {Yoshioka}, \citenamefont {Kuwata-Gonokami}, \citenamefont {Ono},
  \citenamefont {Kumigashira}, \citenamefont {B\"ohmer}, \citenamefont {Hardy},
  \citenamefont {Wolf}, \citenamefont {Meingast}, \citenamefont {L\"ohneysen},
  \citenamefont {Ikeda},\ and\ \citenamefont {Ishizaka}}]{shimojima14}%
  \BibitemOpen
  \bibfield  {author} {\bibinfo {author} {\bibfnamefont {T.}~\bibnamefont
  {Shimojima}}, \bibinfo {author} {\bibfnamefont {Y.}~\bibnamefont {Suzuki}},
  \bibinfo {author} {\bibfnamefont {T.}~\bibnamefont {Sonobe}}, \bibinfo
  {author} {\bibfnamefont {A.}~\bibnamefont {Nakamura}}, \bibinfo {author}
  {\bibfnamefont {M.}~\bibnamefont {Sakano}}, \bibinfo {author} {\bibfnamefont
  {J.}~\bibnamefont {Omachi}}, \bibinfo {author} {\bibfnamefont
  {K.}~\bibnamefont {Yoshioka}}, \bibinfo {author} {\bibfnamefont
  {M.}~\bibnamefont {Kuwata-Gonokami}}, \bibinfo {author} {\bibfnamefont
  {K.}~\bibnamefont {Ono}}, \bibinfo {author} {\bibfnamefont {H.}~\bibnamefont
  {Kumigashira}}, \bibinfo {author} {\bibfnamefont {A.~E.}\ \bibnamefont
  {B\"ohmer}}, \bibinfo {author} {\bibfnamefont {F.}~\bibnamefont {Hardy}},
  \bibinfo {author} {\bibfnamefont {T.}~\bibnamefont {Wolf}}, \bibinfo {author}
  {\bibfnamefont {C.}~\bibnamefont {Meingast}}, \bibinfo {author}
  {\bibfnamefont {H.~v.}\ \bibnamefont {L\"ohneysen}}, \bibinfo {author}
  {\bibfnamefont {H.}~\bibnamefont {Ikeda}}, \ and\ \bibinfo {author}
  {\bibfnamefont {K.}~\bibnamefont {Ishizaka}},\ }\bibfield  {title} {\enquote
  {\bibinfo {title} {{Lifting of xz/yz orbital degeneracy at the structural
  transition in detwinned FeSe}},}\ }\href {\doibase
  10.1103/PhysRevB.90.121111} {\bibfield  {journal} {\bibinfo  {journal} {Phys.
  Rev. B}\ }\textbf {\bibinfo {volume} {90}},\ \bibinfo {pages} {121111}
  (\bibinfo {year} {2014})}\BibitemShut {NoStop}%
\bibitem [{\citenamefont {B\"ohmer}\ \emph {et~al.}(2015)\citenamefont
  {B\"ohmer}, \citenamefont {Arai}, \citenamefont {Hardy}, \citenamefont
  {Hattori}, \citenamefont {Iye}, \citenamefont {Wolf}, \citenamefont
  {L\"ohneysen}, \citenamefont {Ishida},\ and\ \citenamefont
  {Meingast}}]{bohmer15}%
  \BibitemOpen
  \bibfield  {author} {\bibinfo {author} {\bibfnamefont {A.~E.}\ \bibnamefont
  {B\"ohmer}}, \bibinfo {author} {\bibfnamefont {T.}~\bibnamefont {Arai}},
  \bibinfo {author} {\bibfnamefont {F.}~\bibnamefont {Hardy}}, \bibinfo
  {author} {\bibfnamefont {T.}~\bibnamefont {Hattori}}, \bibinfo {author}
  {\bibfnamefont {T.}~\bibnamefont {Iye}}, \bibinfo {author} {\bibfnamefont
  {T.}~\bibnamefont {Wolf}}, \bibinfo {author} {\bibfnamefont {H.~v.}\
  \bibnamefont {L\"ohneysen}}, \bibinfo {author} {\bibfnamefont
  {K.}~\bibnamefont {Ishida}}, \ and\ \bibinfo {author} {\bibfnamefont
  {C.}~\bibnamefont {Meingast}},\ }\bibfield  {title} {\enquote {\bibinfo
  {title} {{Origin of the Tetragonal-to-Orthorhombic Phase Transition in FeSe:
  A Combined Thermodynamic and NMR Study of Nematicity}},}\ }\href {\doibase
  10.1103/PhysRevLett.114.027001} {\bibfield  {journal} {\bibinfo  {journal}
  {Phys. Rev. Lett.}\ }\textbf {\bibinfo {volume} {114}},\ \bibinfo {pages}
  {027001} (\bibinfo {year} {2015})}\BibitemShut {NoStop}%
\bibitem [{\citenamefont {Rahn}\ \emph {et~al.}(2015)\citenamefont {Rahn},
  \citenamefont {Ewings}, \citenamefont {Sedlmaier}, \citenamefont {Clarke},\
  and\ \citenamefont {Boothroyd}}]{rahn15}%
  \BibitemOpen
  \bibfield  {author} {\bibinfo {author} {\bibfnamefont {M.~C.}\ \bibnamefont
  {Rahn}}, \bibinfo {author} {\bibfnamefont {R.~A.}\ \bibnamefont {Ewings}},
  \bibinfo {author} {\bibfnamefont {S.~J.}\ \bibnamefont {Sedlmaier}}, \bibinfo
  {author} {\bibfnamefont {S.~J.}\ \bibnamefont {Clarke}}, \ and\ \bibinfo
  {author} {\bibfnamefont {A.~T.}\ \bibnamefont {Boothroyd}},\ }\bibfield
  {title} {\enquote {\bibinfo {title} {{Strong $(\ensuremath{\pi},0)$ spin
  fluctuations in $\ensuremath{\beta}-\mathrm{FeSe}$ observed by neutron
  spectroscopy}},}\ }\href {\doibase 10.1103/PhysRevB.91.180501} {\bibfield
  {journal} {\bibinfo  {journal} {Phys. Rev. B}\ }\textbf {\bibinfo {volume}
  {91}},\ \bibinfo {pages} {180501} (\bibinfo {year} {2015})}\BibitemShut
  {NoStop}%
\bibitem [{\citenamefont {Wang}\ \emph {et~al.}(2016)\citenamefont {Wang},
  \citenamefont {Shen}, \citenamefont {Pan}, \citenamefont {Hao}, \citenamefont
  {Ma}, \citenamefont {Zhou}, \citenamefont {Steffens}, \citenamefont
  {Schmalzl}, \citenamefont {Forrest}, \citenamefont {Abdel-Hafiez},
  \citenamefont {Chareev}, \citenamefont {Vasiliev}, \citenamefont {Bourges},
  \citenamefont {Sidis}, \citenamefont {Cao},\ and\ \citenamefont
  {Zhao}}]{wang15a}%
  \BibitemOpen
  \bibfield  {author} {\bibinfo {author} {\bibfnamefont {Q.}~\bibnamefont
  {Wang}}, \bibinfo {author} {\bibfnamefont {Y.}~\bibnamefont {Shen}}, \bibinfo
  {author} {\bibfnamefont {B.}~\bibnamefont {Pan}}, \bibinfo {author}
  {\bibfnamefont {Y.}~\bibnamefont {Hao}}, \bibinfo {author} {\bibfnamefont
  {M.}~\bibnamefont {Ma}}, \bibinfo {author} {\bibfnamefont {F.}~\bibnamefont
  {Zhou}}, \bibinfo {author} {\bibfnamefont {P.}~\bibnamefont {Steffens}},
  \bibinfo {author} {\bibfnamefont {K.}~\bibnamefont {Schmalzl}}, \bibinfo
  {author} {\bibfnamefont {T.~R.}\ \bibnamefont {Forrest}}, \bibinfo {author}
  {\bibfnamefont {M.}~\bibnamefont {Abdel-Hafiez}}, \bibinfo {author}
  {\bibfnamefont {D.~A.}\ \bibnamefont {Chareev}}, \bibinfo {author}
  {\bibfnamefont {A.~N.}\ \bibnamefont {Vasiliev}}, \bibinfo {author}
  {\bibfnamefont {P.}~\bibnamefont {Bourges}}, \bibinfo {author} {\bibfnamefont
  {Y.}~\bibnamefont {Sidis}}, \bibinfo {author} {\bibfnamefont
  {H.}~\bibnamefont {Cao}}, \ and\ \bibinfo {author} {\bibfnamefont
  {J.}~\bibnamefont {Zhao}},\ }\bibfield  {title} {\enquote {\bibinfo {title}
  {{Strong Interplay between Stripe Spin Fluctuations, Nematicity and
  Superconductivity in FeSe}},}\ }\href {\doibase 10.1038/nmat4492} {\bibfield
  {journal} {\bibinfo  {journal} {Nature Mater.}\ }\textbf {\bibinfo {volume}
  {15}},\ \bibinfo {pages} {150--163} (\bibinfo {year} {2016})}\BibitemShut
  {NoStop}%
\bibitem [{\citenamefont {Watson}\ \emph {et~al.}(2015)\citenamefont {Watson},
  \citenamefont {Kim}, \citenamefont {Haghighirad}, \citenamefont {Davies},
  \citenamefont {McCollam}, \citenamefont {Narayanan}, \citenamefont {Blake},
  \citenamefont {Chen}, \citenamefont {Ghannadzadeh}, \citenamefont
  {Schofield}, \citenamefont {Hoesch}, \citenamefont {Meingast}, \citenamefont
  {Wolf},\ and\ \citenamefont {Coldea}}]{watson15}%
  \BibitemOpen
  \bibfield  {author} {\bibinfo {author} {\bibfnamefont {M.~D.}\ \bibnamefont
  {Watson}}, \bibinfo {author} {\bibfnamefont {T.~K.}\ \bibnamefont {Kim}},
  \bibinfo {author} {\bibfnamefont {A.~A.}\ \bibnamefont {Haghighirad}},
  \bibinfo {author} {\bibfnamefont {N.~R.}\ \bibnamefont {Davies}}, \bibinfo
  {author} {\bibfnamefont {A.}~\bibnamefont {McCollam}}, \bibinfo {author}
  {\bibfnamefont {A.}~\bibnamefont {Narayanan}}, \bibinfo {author}
  {\bibfnamefont {S.~F.}\ \bibnamefont {Blake}}, \bibinfo {author}
  {\bibfnamefont {Y.~L.}\ \bibnamefont {Chen}}, \bibinfo {author}
  {\bibfnamefont {S.}~\bibnamefont {Ghannadzadeh}}, \bibinfo {author}
  {\bibfnamefont {A.~J.}\ \bibnamefont {Schofield}}, \bibinfo {author}
  {\bibfnamefont {M.}~\bibnamefont {Hoesch}}, \bibinfo {author} {\bibfnamefont
  {C.}~\bibnamefont {Meingast}}, \bibinfo {author} {\bibfnamefont
  {T.}~\bibnamefont {Wolf}}, \ and\ \bibinfo {author} {\bibfnamefont {A.~I.}\
  \bibnamefont {Coldea}},\ }\bibfield  {title} {\enquote {\bibinfo {title}
  {{Emergence of the nematic electronic state in FeSe}},}\ }\href {\doibase
  10.1103/PhysRevB.91.155106} {\bibfield  {journal} {\bibinfo  {journal} {Phys.
  Rev. B}\ }\textbf {\bibinfo {volume} {91}},\ \bibinfo {pages} {155106}
  (\bibinfo {year} {2015})}\BibitemShut {NoStop}%
\bibitem [{\citenamefont {Mukherjee}\ \emph {et~al.}(2015)\citenamefont
  {Mukherjee}, \citenamefont {Kreisel}, \citenamefont {Hirschfeld},\ and\
  \citenamefont {Andersen}}]{mukherjee15}%
  \BibitemOpen
  \bibfield  {author} {\bibinfo {author} {\bibfnamefont {S.}~\bibnamefont
  {Mukherjee}}, \bibinfo {author} {\bibfnamefont {A.}~\bibnamefont {Kreisel}},
  \bibinfo {author} {\bibfnamefont {P.~J.}\ \bibnamefont {Hirschfeld}}, \ and\
  \bibinfo {author} {\bibfnamefont {B.~M.}\ \bibnamefont {Andersen}},\
  }\bibfield  {title} {\enquote {\bibinfo {title} {{Model of Electronic
  Structure and Superconductivity in Orbitally Ordered FeSe}},}\ }\href
  {\doibase 10.1103/PhysRevLett.115.026402} {\bibfield  {journal} {\bibinfo
  {journal} {Phys. Rev. Lett.}\ }\textbf {\bibinfo {volume} {115}},\ \bibinfo
  {pages} {026402} (\bibinfo {year} {2015})}\BibitemShut {NoStop}%
\bibitem [{\citenamefont {Glasbrenner}\ \emph {et~al.}(2015)\citenamefont
  {Glasbrenner}, \citenamefont {Mazin}, \citenamefont {Jeschke}, \citenamefont
  {Hirschfeld}, \citenamefont {Fernandes},\ and\ \citenamefont
  {Valenti}}]{glasbrenner15}%
  \BibitemOpen
  \bibfield  {author} {\bibinfo {author} {\bibfnamefont {J.~K.}\ \bibnamefont
  {Glasbrenner}}, \bibinfo {author} {\bibfnamefont {I.~I.}\ \bibnamefont
  {Mazin}}, \bibinfo {author} {\bibfnamefont {H.~O.}\ \bibnamefont {Jeschke}},
  \bibinfo {author} {\bibfnamefont {P.~J.}\ \bibnamefont {Hirschfeld}},
  \bibinfo {author} {\bibfnamefont {R.~M.}\ \bibnamefont {Fernandes}}, \ and\
  \bibinfo {author} {\bibfnamefont {R.}~\bibnamefont {Valenti}},\ }\bibfield
  {title} {\enquote {\bibinfo {title} {{Effect of magnetic frustration on
  nematicity and superconductivity in iron chalcogenides}},}\ }\href {\doibase
  10.1038/nphys3434} {\bibfield  {journal} {\bibinfo  {journal} {Nature Phys.}\
  }\textbf {\bibinfo {volume} {11}},\ \bibinfo {pages} {953--958} (\bibinfo
  {year} {2015})}\BibitemShut {NoStop}%
\bibitem [{\citenamefont {R\"o\ss{}ler}\ \emph {et~al.}(2015)\citenamefont
  {R\"o\ss{}ler}, \citenamefont {Koz}, \citenamefont {Jiao}, \citenamefont
  {R\"o\ss{}ler}, \citenamefont {Steglich}, \citenamefont {Schwarz},\ and\
  \citenamefont {Wirth}}]{rossler15}%
  \BibitemOpen
  \bibfield  {author} {\bibinfo {author} {\bibfnamefont {S.}~\bibnamefont
  {R\"o\ss{}ler}}, \bibinfo {author} {\bibfnamefont {C.}~\bibnamefont {Koz}},
  \bibinfo {author} {\bibfnamefont {L.}~\bibnamefont {Jiao}}, \bibinfo {author}
  {\bibfnamefont {U.~K.}\ \bibnamefont {R\"o\ss{}ler}}, \bibinfo {author}
  {\bibfnamefont {F.}~\bibnamefont {Steglich}}, \bibinfo {author}
  {\bibfnamefont {U.}~\bibnamefont {Schwarz}}, \ and\ \bibinfo {author}
  {\bibfnamefont {S.}~\bibnamefont {Wirth}},\ }\bibfield  {title} {\enquote
  {\bibinfo {title} {{Emergence of an incipient ordering mode in FeSe}},}\
  }\href {\doibase 10.1103/PhysRevB.92.060505} {\bibfield  {journal} {\bibinfo
  {journal} {Phys. Rev. B}\ }\textbf {\bibinfo {volume} {92}},\ \bibinfo
  {pages} {060505} (\bibinfo {year} {2015})}\BibitemShut {NoStop}%
\bibitem [{\citenamefont {Margadonna}\ \emph {et~al.}(2009)\citenamefont
  {Margadonna}, \citenamefont {Takabayashi}, \citenamefont {Ohishi},
  \citenamefont {Mizuguchi}, \citenamefont {Takano}, \citenamefont {Kagayama},
  \citenamefont {Nakagawa}, \citenamefont {Takata},\ and\ \citenamefont
  {Prassides}}]{margadonna09}%
  \BibitemOpen
  \bibfield  {author} {\bibinfo {author} {\bibfnamefont {S.}~\bibnamefont
  {Margadonna}}, \bibinfo {author} {\bibfnamefont {Y.}~\bibnamefont
  {Takabayashi}}, \bibinfo {author} {\bibfnamefont {Y.}~\bibnamefont {Ohishi}},
  \bibinfo {author} {\bibfnamefont {Y.}~\bibnamefont {Mizuguchi}}, \bibinfo
  {author} {\bibfnamefont {Y.}~\bibnamefont {Takano}}, \bibinfo {author}
  {\bibfnamefont {T.}~\bibnamefont {Kagayama}}, \bibinfo {author}
  {\bibfnamefont {T.}~\bibnamefont {Nakagawa}}, \bibinfo {author}
  {\bibfnamefont {M.}~\bibnamefont {Takata}}, \ and\ \bibinfo {author}
  {\bibfnamefont {K.}~\bibnamefont {Prassides}},\ }\bibfield  {title} {\enquote
  {\bibinfo {title} {{Pressure evolution of the low-temperature crystal
  structure and bonding of the superconductor FeSe ($T_c$=37 K)}},}\ }\href
  {\doibase 10.1103/PhysRevB.80.064506} {\bibfield  {journal} {\bibinfo
  {journal} {Phys. Rev. B}\ }\textbf {\bibinfo {volume} {80}},\ \bibinfo
  {pages} {064506} (\bibinfo {year} {2009})}\BibitemShut {NoStop}%
\bibitem [{\citenamefont {Medvedev}\ \emph {et~al.}(2009)\citenamefont
  {Medvedev}, \citenamefont {McQueen}, \citenamefont {Troyan}, \citenamefont
  {Palasyuk}, \citenamefont {Eremets}, \citenamefont {Cava}, \citenamefont
  {Naghavi}, \citenamefont {Casper}, \citenamefont {Ksenofontov}, \citenamefont
  {Wortmann},\ and\ \citenamefont {Felser}}]{medvedev09}%
  \BibitemOpen
  \bibfield  {author} {\bibinfo {author} {\bibfnamefont {S.}~\bibnamefont
  {Medvedev}}, \bibinfo {author} {\bibfnamefont {T.~M.}\ \bibnamefont
  {McQueen}}, \bibinfo {author} {\bibfnamefont {I.~A.}\ \bibnamefont {Troyan}},
  \bibinfo {author} {\bibfnamefont {T.}~\bibnamefont {Palasyuk}}, \bibinfo
  {author} {\bibfnamefont {M.~I.}\ \bibnamefont {Eremets}}, \bibinfo {author}
  {\bibfnamefont {R.~J.}\ \bibnamefont {Cava}}, \bibinfo {author}
  {\bibfnamefont {S.}~\bibnamefont {Naghavi}}, \bibinfo {author} {\bibfnamefont
  {F.}~\bibnamefont {Casper}}, \bibinfo {author} {\bibfnamefont
  {V.}~\bibnamefont {Ksenofontov}}, \bibinfo {author} {\bibfnamefont
  {G.}~\bibnamefont {Wortmann}}, \ and\ \bibinfo {author} {\bibfnamefont
  {C.}~\bibnamefont {Felser}},\ }\bibfield  {title} {\enquote {\bibinfo {title}
  {{Electronic and magnetic phase diagram of {$\beta$-Fe$_{1.01}$Se} with
  superconductivity at {36.7 K} under pressure}},}\ }\href {\doibase
  10.1038/nmat2491} {\bibfield  {journal} {\bibinfo  {journal} {Nature Mater.}\
  }\textbf {\bibinfo {volume} {8}},\ \bibinfo {pages} {630--633} (\bibinfo
  {year} {2009})}\BibitemShut {NoStop}%
\bibitem [{\citenamefont {Xiang}\ \emph {et~al.}(2012)\citenamefont {Xiang},
  \citenamefont {Wang}, \citenamefont {Wang}, \citenamefont {Wang},\ and\
  \citenamefont {Lee}}]{xiang12}%
  \BibitemOpen
  \bibfield  {author} {\bibinfo {author} {\bibfnamefont {Y.-Y.}\ \bibnamefont
  {Xiang}}, \bibinfo {author} {\bibfnamefont {F.}~\bibnamefont {Wang}},
  \bibinfo {author} {\bibfnamefont {D.}~\bibnamefont {Wang}}, \bibinfo {author}
  {\bibfnamefont {Q.-H.}\ \bibnamefont {Wang}}, \ and\ \bibinfo {author}
  {\bibfnamefont {D.-H.}\ \bibnamefont {Lee}},\ }\bibfield  {title} {\enquote
  {\bibinfo {title} {{High-temperature superconductivity at the
  FeSe/SrTiO$_{3}$ interface}},}\ }\href {\doibase 10.1103/PhysRevB.86.134508}
  {\bibfield  {journal} {\bibinfo  {journal} {Phys. Rev. B}\ }\textbf {\bibinfo
  {volume} {86}},\ \bibinfo {pages} {134508} (\bibinfo {year}
  {2012})}\BibitemShut {NoStop}%
\bibitem [{\citenamefont {He}\ \emph {et~al.}(2013)\citenamefont {He},
  \citenamefont {He}, \citenamefont {Zhang}, \citenamefont {Zhao},
  \citenamefont {Liu}, \citenamefont {Liu}, \citenamefont {Mou}, \citenamefont
  {Ou}, \citenamefont {Wang}, \citenamefont {Li}, \citenamefont {Wang},
  \citenamefont {Peng}, \citenamefont {Liu}, \citenamefont {Chen},
  \citenamefont {Yu}, \citenamefont {Liu}, \citenamefont {Dong}, \citenamefont
  {Zhang}, \citenamefont {Chen}, \citenamefont {Xu}, \citenamefont {Chen},
  \citenamefont {Ma}, \citenamefont {Xue},\ and\ \citenamefont {Zhou}}]{he13}%
  \BibitemOpen
  \bibfield  {author} {\bibinfo {author} {\bibfnamefont {S.}~\bibnamefont
  {He}}, \bibinfo {author} {\bibfnamefont {J.}~\bibnamefont {He}}, \bibinfo
  {author} {\bibfnamefont {W.}~\bibnamefont {Zhang}}, \bibinfo {author}
  {\bibfnamefont {L.}~\bibnamefont {Zhao}}, \bibinfo {author} {\bibfnamefont
  {D.}~\bibnamefont {Liu}}, \bibinfo {author} {\bibfnamefont {X.}~\bibnamefont
  {Liu}}, \bibinfo {author} {\bibfnamefont {D.}~\bibnamefont {Mou}}, \bibinfo
  {author} {\bibfnamefont {Y.-B.}\ \bibnamefont {Ou}}, \bibinfo {author}
  {\bibfnamefont {Q.-Y.}\ \bibnamefont {Wang}}, \bibinfo {author}
  {\bibfnamefont {Z.}~\bibnamefont {Li}}, \bibinfo {author} {\bibfnamefont
  {L.}~\bibnamefont {Wang}}, \bibinfo {author} {\bibfnamefont {Y.}~\bibnamefont
  {Peng}}, \bibinfo {author} {\bibfnamefont {Y.}~\bibnamefont {Liu}}, \bibinfo
  {author} {\bibfnamefont {C.}~\bibnamefont {Chen}}, \bibinfo {author}
  {\bibfnamefont {L.}~\bibnamefont {Yu}}, \bibinfo {author} {\bibfnamefont
  {G.}~\bibnamefont {Liu}}, \bibinfo {author} {\bibfnamefont {X.}~\bibnamefont
  {Dong}}, \bibinfo {author} {\bibfnamefont {J.}~\bibnamefont {Zhang}},
  \bibinfo {author} {\bibfnamefont {C.}~\bibnamefont {Chen}}, \bibinfo {author}
  {\bibfnamefont {Z.}~\bibnamefont {Xu}}, \bibinfo {author} {\bibfnamefont
  {X.}~\bibnamefont {Chen}}, \bibinfo {author} {\bibfnamefont {X.}~\bibnamefont
  {Ma}}, \bibinfo {author} {\bibfnamefont {Q.}~\bibnamefont {Xue}}, \ and\
  \bibinfo {author} {\bibfnamefont {X.~J.}\ \bibnamefont {Zhou}},\ }\bibfield
  {title} {\enquote {\bibinfo {title} {Phase diagram and electronic indication
  of high-temperature superconductivity at {65 K} in single-layer {FeSe}
  films},}\ }\href {\doibase 10.1038/nmat3648} {\bibfield  {journal} {\bibinfo
  {journal} {Nature Mater.}\ }\textbf {\bibinfo {volume} {12}},\ \bibinfo
  {pages} {605--610} (\bibinfo {year} {2013})}\BibitemShut {NoStop}%
\bibitem [{\citenamefont {Ge}\ \emph {et~al.}(2015)\citenamefont {Ge},
  \citenamefont {Liu}, \citenamefont {Liu}, \citenamefont {Gao}, \citenamefont
  {Qian}, \citenamefont {Xue}, \citenamefont {Liu},\ and\ \citenamefont
  {Jia}}]{ge15}%
  \BibitemOpen
  \bibfield  {author} {\bibinfo {author} {\bibfnamefont {J.-F.}\ \bibnamefont
  {Ge}}, \bibinfo {author} {\bibfnamefont {Z.-L.}\ \bibnamefont {Liu}},
  \bibinfo {author} {\bibfnamefont {C.}~\bibnamefont {Liu}}, \bibinfo {author}
  {\bibfnamefont {C.-L.}\ \bibnamefont {Gao}}, \bibinfo {author} {\bibfnamefont
  {D.}~\bibnamefont {Qian}}, \bibinfo {author} {\bibfnamefont {Q.-K.}\
  \bibnamefont {Xue}}, \bibinfo {author} {\bibfnamefont {Y.}~\bibnamefont
  {Liu}}, \ and\ \bibinfo {author} {\bibfnamefont {J.-F.}\ \bibnamefont
  {Jia}},\ }\bibfield  {title} {\enquote {\bibinfo {title} {{Superconductivity
  above 100 K in single-layer FeSe films on doped SrTiO$_3$}},}\ }\href
  {\doibase 10.1038/nmat4153} {\bibfield  {journal} {\bibinfo  {journal}
  {Nature Mater.}\ }\textbf {\bibinfo {volume} {14}},\ \bibinfo {pages}
  {285--289} (\bibinfo {year} {2015})}\BibitemShut {NoStop}%
\bibitem [{\citenamefont {Maletz}\ \emph {et~al.}(2014)\citenamefont {Maletz},
  \citenamefont {Zabolotnyy}, \citenamefont {Evtushinsky}, \citenamefont
  {Thirupathaiah}, \citenamefont {Wolter}, \citenamefont {Harnagea},
  \citenamefont {Yaresko}, \citenamefont {Vasiliev}, \citenamefont {Chareev},
  \citenamefont {B\"ohmer}, \citenamefont {Hardy}, \citenamefont {Wolf},
  \citenamefont {Meingast}, \citenamefont {Rienks}, \citenamefont {B\"uchner},\
  and\ \citenamefont {Borisenko}}]{maletz14}%
  \BibitemOpen
  \bibfield  {author} {\bibinfo {author} {\bibfnamefont {J.}~\bibnamefont
  {Maletz}}, \bibinfo {author} {\bibfnamefont {V.~B.}\ \bibnamefont
  {Zabolotnyy}}, \bibinfo {author} {\bibfnamefont {D.~V.}\ \bibnamefont
  {Evtushinsky}}, \bibinfo {author} {\bibfnamefont {S.}~\bibnamefont
  {Thirupathaiah}}, \bibinfo {author} {\bibfnamefont {A.~U.~B.}\ \bibnamefont
  {Wolter}}, \bibinfo {author} {\bibfnamefont {L.}~\bibnamefont {Harnagea}},
  \bibinfo {author} {\bibfnamefont {A.~N.}\ \bibnamefont {Yaresko}}, \bibinfo
  {author} {\bibfnamefont {A.~N.}\ \bibnamefont {Vasiliev}}, \bibinfo {author}
  {\bibfnamefont {D.~A.}\ \bibnamefont {Chareev}}, \bibinfo {author}
  {\bibfnamefont {A.~E.}\ \bibnamefont {B\"ohmer}}, \bibinfo {author}
  {\bibfnamefont {F.}~\bibnamefont {Hardy}}, \bibinfo {author} {\bibfnamefont
  {T.}~\bibnamefont {Wolf}}, \bibinfo {author} {\bibfnamefont {C.}~\bibnamefont
  {Meingast}}, \bibinfo {author} {\bibfnamefont {E.~D.~L.}\ \bibnamefont
  {Rienks}}, \bibinfo {author} {\bibfnamefont {B.}~\bibnamefont {B\"uchner}}, \
  and\ \bibinfo {author} {\bibfnamefont {S.~V.}\ \bibnamefont {Borisenko}},\
  }\bibfield  {title} {\enquote {\bibinfo {title} {{Unusual band
  renormalization in the simplest iron-based superconductor FeSe$_{1-x}$}},}\
  }\href {\doibase 10.1103/PhysRevB.89.220506} {\bibfield  {journal} {\bibinfo
  {journal} {Phys. Rev. B}\ }\textbf {\bibinfo {volume} {89}},\ \bibinfo
  {pages} {220506} (\bibinfo {year} {2014})}\BibitemShut {NoStop}%
\bibitem [{\citenamefont {Zhang}\ \emph {et~al.}(2015)\citenamefont {Zhang},
  \citenamefont {Qian}, \citenamefont {Richard}, \citenamefont {Wang},
  \citenamefont {Miao}, \citenamefont {Lv}, \citenamefont {Fu}, \citenamefont
  {Wolf}, \citenamefont {Meingast}, \citenamefont {Wu}, \citenamefont {Wang},
  \citenamefont {Hu},\ and\ \citenamefont {Ding}}]{zhang15}%
  \BibitemOpen
  \bibfield  {author} {\bibinfo {author} {\bibfnamefont {P.}~\bibnamefont
  {Zhang}}, \bibinfo {author} {\bibfnamefont {T.}~\bibnamefont {Qian}},
  \bibinfo {author} {\bibfnamefont {P.}~\bibnamefont {Richard}}, \bibinfo
  {author} {\bibfnamefont {X.~P.}\ \bibnamefont {Wang}}, \bibinfo {author}
  {\bibfnamefont {H.}~\bibnamefont {Miao}}, \bibinfo {author} {\bibfnamefont
  {B.~Q.}\ \bibnamefont {Lv}}, \bibinfo {author} {\bibfnamefont {B.~B.}\
  \bibnamefont {Fu}}, \bibinfo {author} {\bibfnamefont {T.}~\bibnamefont
  {Wolf}}, \bibinfo {author} {\bibfnamefont {C.}~\bibnamefont {Meingast}},
  \bibinfo {author} {\bibfnamefont {X.~X.}\ \bibnamefont {Wu}}, \bibinfo
  {author} {\bibfnamefont {Z.~Q.}\ \bibnamefont {Wang}}, \bibinfo {author}
  {\bibfnamefont {J.~P.}\ \bibnamefont {Hu}}, \ and\ \bibinfo {author}
  {\bibfnamefont {H.}~\bibnamefont {Ding}},\ }\bibfield  {title} {\enquote
  {\bibinfo {title} {{Observation of two distinct ${d}_{xz}$/${d}_{yz}$ band
  splittings in FeSe}},}\ }\href {\doibase 10.1103/PhysRevB.91.214503}
  {\bibfield  {journal} {\bibinfo  {journal} {Phys. Rev. B}\ }\textbf {\bibinfo
  {volume} {91}},\ \bibinfo {pages} {214503} (\bibinfo {year}
  {2015})}\BibitemShut {NoStop}%
\bibitem [{\citenamefont {Wen}\ \emph {et~al.}(2012)\citenamefont {Wen},
  \citenamefont {Wang}, \citenamefont {Chang}, \citenamefont {Luo},
  \citenamefont {Shen}, \citenamefont {Liu}, \citenamefont {Sun}, \citenamefont
  {Wang},\ and\ \citenamefont {Wu}}]{wen12b}%
  \BibitemOpen
  \bibfield  {author} {\bibinfo {author} {\bibfnamefont {Y.-C.}\ \bibnamefont
  {Wen}}, \bibinfo {author} {\bibfnamefont {K.-J.}\ \bibnamefont {Wang}},
  \bibinfo {author} {\bibfnamefont {H.-H.}\ \bibnamefont {Chang}}, \bibinfo
  {author} {\bibfnamefont {J.-Y.}\ \bibnamefont {Luo}}, \bibinfo {author}
  {\bibfnamefont {C.-C.}\ \bibnamefont {Shen}}, \bibinfo {author}
  {\bibfnamefont {H.-L.}\ \bibnamefont {Liu}}, \bibinfo {author} {\bibfnamefont
  {C.-K.}\ \bibnamefont {Sun}}, \bibinfo {author} {\bibfnamefont {M.-J.}\
  \bibnamefont {Wang}}, \ and\ \bibinfo {author} {\bibfnamefont {M.-K.}\
  \bibnamefont {Wu}},\ }\bibfield  {title} {\enquote {\bibinfo {title} {{Gap
  Opening and Orbital Modification of Superconducting {FeSe} above the
  Structural Distortion}},}\ }\href {\doibase 10.1103/PhysRevLett.108.267002}
  {\bibfield  {journal} {\bibinfo  {journal} {Phys. Rev. Lett.}\ }\textbf
  {\bibinfo {volume} {108}},\ \bibinfo {pages} {267002} (\bibinfo {year}
  {2012})}\BibitemShut {NoStop}%
\bibitem [{\citenamefont {Baek}\ \emph {et~al.}(2015)\citenamefont {Baek},
  \citenamefont {Efremov}, \citenamefont {Ok}, \citenamefont {Kim},
  \citenamefont {van~den Brink},\ and\ \citenamefont {B\"uchner}}]{baek15}%
  \BibitemOpen
  \bibfield  {author} {\bibinfo {author} {\bibfnamefont {S.-H.}\ \bibnamefont
  {Baek}}, \bibinfo {author} {\bibfnamefont {D.~V.}\ \bibnamefont {Efremov}},
  \bibinfo {author} {\bibfnamefont {J.~M.}\ \bibnamefont {Ok}}, \bibinfo
  {author} {\bibfnamefont {J.~S.}\ \bibnamefont {Kim}}, \bibinfo {author}
  {\bibfnamefont {J.}~\bibnamefont {van~den Brink}}, \ and\ \bibinfo {author}
  {\bibfnamefont {B.}~\bibnamefont {B\"uchner}},\ }\bibfield  {title} {\enquote
  {\bibinfo {title} {{Orbital-driven nematicity in FeSe}},}\ }\href {\doibase
  10.1038/nmat4138} {\bibfield  {journal} {\bibinfo  {journal} {Nature Mater.}\
  }\textbf {\bibinfo {volume} {14}},\ \bibinfo {pages} {210--214} (\bibinfo
  {year} {2015})}\BibitemShut {NoStop}%
\bibitem [{\citenamefont {Yamakawa}\ \emph {et~al.}()\citenamefont {Yamakawa},
  \citenamefont {Onari},\ and\ \citenamefont {Kontani}}]{yamakawa15}%
  \BibitemOpen
  \bibfield  {author} {\bibinfo {author} {\bibfnamefont {Y.}~\bibnamefont
  {Yamakawa}}, \bibinfo {author} {\bibfnamefont {S.}~\bibnamefont {Onari}}, \
  and\ \bibinfo {author} {\bibfnamefont {H.}~\bibnamefont {Kontani}},\
  }\bibfield  {title} {\enquote {\bibinfo {title} {{Nematicity and Magnetism in
  FeSe and other Families of Fe-based Superconductors}},}\ }\href@noop {} {\
  }\bibinfo {note} {{arXiv:1509.01161}}\BibitemShut {NoStop}%
\bibitem [{\citenamefont {Onari}\ \emph {et~al.}()\citenamefont {Onari},
  \citenamefont {Yamakawa},\ and\ \citenamefont {Kontani}}]{onari15}%
  \BibitemOpen
  \bibfield  {author} {\bibinfo {author} {\bibfnamefont {S.}~\bibnamefont
  {Onari}}, \bibinfo {author} {\bibfnamefont {Y.}~\bibnamefont {Yamakawa}}, \
  and\ \bibinfo {author} {\bibfnamefont {H.}~\bibnamefont {Kontani}},\
  }\bibfield  {title} {\enquote {\bibinfo {title} {{Sign-Reversing Orbital
  Polarization in the Nematic Phase of FeSe Driven by Aslamazov-Larkin
  Processes}},}\ }\href@noop {} {\ }\bibinfo {note}
  {{arXiv:1509.01172}}\BibitemShut {NoStop}%
\bibitem [{\citenamefont {B\"ohmer}\ and\ \citenamefont
  {Meingast}(2015)}]{bohmer15a}%
  \BibitemOpen
  \bibfield  {author} {\bibinfo {author} {\bibfnamefont {A.~E.}\ \bibnamefont
  {B\"ohmer}}\ and\ \bibinfo {author} {\bibfnamefont {C.}~\bibnamefont
  {Meingast}},\ }\bibfield  {title} {\enquote {\bibinfo {title} {{Electronic
  nematic susceptibility of iron-based superconductors}},}\ }\href {\doibase
  http://dx.doi.org/10.1016/j.crhy.2015.07.001} {\bibfield  {journal} {\bibinfo
   {journal} {C. R. Physique}\ }\textbf {\bibinfo {volume} {17}},\ \bibinfo
  {pages} {90} (\bibinfo {year} {2015})}\BibitemShut {NoStop}%
\bibitem [{\citenamefont {Kasahara}\ \emph {et~al.}(2012)\citenamefont
  {Kasahara}, \citenamefont {Shi}, \citenamefont {Hashimoto}, \citenamefont
  {Tonegawa}, \citenamefont {Mizukami}, \citenamefont {Shibauchi},
  \citenamefont {Sugimoto}, \citenamefont {Fukuda}, \citenamefont {Terashima},
  \citenamefont {Nevidomskyy},\ and\ \citenamefont {Matsuda}}]{kasahara12a}%
  \BibitemOpen
  \bibfield  {author} {\bibinfo {author} {\bibfnamefont {S.}~\bibnamefont
  {Kasahara}}, \bibinfo {author} {\bibfnamefont {H.~J.}\ \bibnamefont {Shi}},
  \bibinfo {author} {\bibfnamefont {K.}~\bibnamefont {Hashimoto}}, \bibinfo
  {author} {\bibfnamefont {S.}~\bibnamefont {Tonegawa}}, \bibinfo {author}
  {\bibfnamefont {Y.}~\bibnamefont {Mizukami}}, \bibinfo {author}
  {\bibfnamefont {T.}~\bibnamefont {Shibauchi}}, \bibinfo {author}
  {\bibfnamefont {K.}~\bibnamefont {Sugimoto}}, \bibinfo {author}
  {\bibfnamefont {T.}~\bibnamefont {Fukuda}}, \bibinfo {author} {\bibfnamefont
  {T.}~\bibnamefont {Terashima}}, \bibinfo {author} {\bibfnamefont {Andriy~H.}\
  \bibnamefont {Nevidomskyy}}, \ and\ \bibinfo {author} {\bibfnamefont
  {Y.}~\bibnamefont {Matsuda}},\ }\bibfield  {title} {\enquote {\bibinfo
  {title} {Electronic nematicity above the structural and superconducting
  transition in {BaFe$_2$(As$_{1-x}$P$_x$)$_2$}},}\ }\href {\doibase
  10.1038/nature11178} {\bibfield  {journal} {\bibinfo  {journal} {Nature}\
  }\textbf {\bibinfo {volume} {486}},\ \bibinfo {pages} {382--385} (\bibinfo
  {year} {2012})}\BibitemShut {NoStop}%
\bibitem [{\citenamefont {B\"ohmer}\ \emph {et~al.}(2013)\citenamefont
  {B\"ohmer}, \citenamefont {Hardy}, \citenamefont {Eilers}, \citenamefont
  {Ernst}, \citenamefont {Adelmann}, \citenamefont {Schweiss}, \citenamefont
  {Wolf},\ and\ \citenamefont {Meingast}}]{bohmer13}%
  \BibitemOpen
  \bibfield  {author} {\bibinfo {author} {\bibfnamefont {A.~E.}\ \bibnamefont
  {B\"ohmer}}, \bibinfo {author} {\bibfnamefont {F.}~\bibnamefont {Hardy}},
  \bibinfo {author} {\bibfnamefont {F.}~\bibnamefont {Eilers}}, \bibinfo
  {author} {\bibfnamefont {D.}~\bibnamefont {Ernst}}, \bibinfo {author}
  {\bibfnamefont {P.}~\bibnamefont {Adelmann}}, \bibinfo {author}
  {\bibfnamefont {P.}~\bibnamefont {Schweiss}}, \bibinfo {author}
  {\bibfnamefont {T.}~\bibnamefont {Wolf}}, \ and\ \bibinfo {author}
  {\bibfnamefont {C.}~\bibnamefont {Meingast}},\ }\bibfield  {title} {\enquote
  {\bibinfo {title} {{Lack of coupling between superconductivity and
  orthorhombic distortion in stoichiometric single-crystalline FeSe}},}\ }\href
  {\doibase 10.1103/PhysRevB.87.180505} {\bibfield  {journal} {\bibinfo
  {journal} {Phys. Rev. B}\ }\textbf {\bibinfo {volume} {87}},\ \bibinfo
  {pages} {180505} (\bibinfo {year} {2013})}\BibitemShut {NoStop}%
\bibitem [{\citenamefont {Kogan}(2002)}]{kogan02}%
  \BibitemOpen
  \bibfield  {author} {\bibinfo {author} {\bibfnamefont {V.~G.}\ \bibnamefont
  {Kogan}},\ }\bibfield  {title} {\enquote {\bibinfo {title} {Macroscopic
  anisotropy in superconductors with anisotropic gaps},}\ }\href {\doibase
  10.1103/PhysRevB.66.020509} {\bibfield  {journal} {\bibinfo  {journal} {Phys.
  Rev. B}\ }\textbf {\bibinfo {volume} {66}},\ \bibinfo {pages} {020509}
  (\bibinfo {year} {2002})}\BibitemShut {NoStop}%
\bibitem [{\citenamefont {Moon}\ and\ \citenamefont {Sachdev}(2012)}]{moon12a}%
  \BibitemOpen
  \bibfield  {author} {\bibinfo {author} {\bibfnamefont {E.-G.}\ \bibnamefont
  {Moon}}\ and\ \bibinfo {author} {\bibfnamefont {S.}~\bibnamefont {Sachdev}},\
  }\bibfield  {title} {\enquote {\bibinfo {title} {{Competition between
  superconductivity and nematic order: Anisotropy of superconducting coherence
  length}},}\ }\href {\doibase 10.1103/PhysRevB.85.184511} {\bibfield
  {journal} {\bibinfo  {journal} {Phys. Rev. B}\ }\textbf {\bibinfo {volume}
  {85}},\ \bibinfo {pages} {184511} (\bibinfo {year} {2012})}\BibitemShut
  {NoStop}%
\bibitem [{\citenamefont {Hung}\ \emph {et~al.}(2012)\citenamefont {Hung},
  \citenamefont {Song}, \citenamefont {Chen}, \citenamefont {Ma}, \citenamefont
  {Xue},\ and\ \citenamefont {Wu}}]{hung12}%
  \BibitemOpen
  \bibfield  {author} {\bibinfo {author} {\bibfnamefont {H.-H.}\ \bibnamefont
  {Hung}}, \bibinfo {author} {\bibfnamefont {C.-L.}\ \bibnamefont {Song}},
  \bibinfo {author} {\bibfnamefont {X.}~\bibnamefont {Chen}}, \bibinfo {author}
  {\bibfnamefont {X.}~\bibnamefont {Ma}}, \bibinfo {author} {\bibfnamefont
  {Q.-k.}\ \bibnamefont {Xue}}, \ and\ \bibinfo {author} {\bibfnamefont
  {C.}~\bibnamefont {Wu}},\ }\bibfield  {title} {\enquote {\bibinfo {title}
  {{Anisotropic vortex lattice structures in the {FeSe} superconductor}},}\
  }\href {\doibase 10.1103/PhysRevB.85.104510} {\bibfield  {journal} {\bibinfo
  {journal} {Phys. Rev. B}\ }\textbf {\bibinfo {volume} {85}},\ \bibinfo
  {pages} {104510} (\bibinfo {year} {2012})}\BibitemShut {NoStop}%
\end{thebibliography}%

\end{document}